\documentclass[review]{elsarticle}
\usepackage{hyperref}
\usepackage{amsmath}
\usepackage{xcolor}
\usepackage{rotating}
\usepackage{natbib}
\usepackage{multirow}
\usepackage[left=1.5cm, right=1.5cm, top=1.785cm, bottom=2.0cm]{geometry}
    
\makeatletter
\def\ps@pprintTitle{%
  \let\@oddhead\@empty
  \let\@evenhead\@empty
  \def\@oddfoot{}%
  \let\@evenfoot\@oddfoot}
\makeatother

\begin{document}

\begin{frontmatter}

\title{Phase field dislocation dynamics prediction of stacking fault widths in Cu–Al Alloys using generalised stacking fault energy from DFT and interatomic potentials}

\author{Arjun Varma R.~$^{a,b}$, Ankosh D. Deshmukh~$^{c,d}$, Sushil Kumar~$^{a,e}$, Vaishali Shah~$^{d}$, M. P. Gururajan~$^{a,f}$}


\address[mymainaddress]{Department of Metallurgical Engineering and Materials Science, Indian Institute of Technology Bombay, Mumbai, India}

\address[mysubaddress]{Institut Jean Lamour, Universit{\'e} de Lorraine, Nancy, France}

\address[myternaryaddress]{Department of Physics, Savitribai Phule Pune University, Pune 411007, India}

\address[mysecondaryaddress]{Department of Scientific Computing, Modeling and Simulation, Savitribai Phule Pune University, Pune 411007, India}

\address[mysubaddress]{{Graduate Institute of Ferrous and Eco Materials Technology, Pohang University of Science and
Technology, Republic of Korea}}

\address[mysubaddress]{{Currrently on sabbatical at Daiichi Sankyo Europe GmbH, 85276 Pfaffenhofen an der Ilm, Germany.}}

\begin{abstract}

The equilibrium width of intrinsic stacking faults is an important
material parameter affecting key deformation phenomena such as
cross-slip, climb, segregation of solutes (hydrogen, for example)
that alters the mobility of extended dislocations, and twinning.
Molecular dynamics (MD) simulations can calculate the equilibrium
stacking fault width (SFW) in metallic systems with good accuracy;
however, studying phenomena such as solute segregation to stacking
faults is far more effectively addressed using a phase field
dislocation dynamics (PFDD) model that incorporates the generalized
stacking fault energy ($\gamma$-surface) calculated from atomistic
simulations. Here, we perform a comparative assessment of
$\gamma$-surfaces generated using MD -- with a classical
embedded-atom potential (ZW) and a machine-learned
Gaussian Approximation Potential (GAP) -- and density functional
theory (DFT), for Cu-Al alloys of varying Al content, and incorporate
them as an energy term in PFDD simulations to obtain the equilibrium
SFW. All three methods reproduce the lattice parameter and elastic
moduli of Cu and Cu-Al alloys in reasonable agreement with each other
and with experimental measurements from literature, but the MD-derived 
$\gamma$-surfaces, for both
ZW and GAP, deviate substantially from DFT at the seven
characteristic points, with the largest discrepancies at $T$ and
$T_1$. This is striking for GAP in particular, since it is trained
directly on DFT data and might have been expected to reproduce the
DFT $\gamma$-surface more closely than a classical potential. When
incorporated into PFDD, the DFT-derived $\gamma$-surfaces
successfully predict an equilibrium SFW that increases with Al
content and agrees well with standalone MD measurements and
experimental values for pure Cu, whereas the MD-derived
$\gamma$-surfaces do not yield an accurate SFW within PFDD, as the
altered generalized stacking fault energy landscape causes the 
dissociating partials to become pinned
soon after separation. Surprisingly, the SFW from standalone MD using
ZW nonetheless agrees well with the DFT-PFDD predictions, indicating
that $\gamma$-surface reliability for PFDD depends on more than its
point-wise accuracy relative to DFT; specifically, even though the 
relative minima are correctly identified by the classical and machine 
learned interatomic potentials, the barrier structure and energetics 
are not captured faithfully. These results show that
DFT-derived $\gamma$-surfaces remain the more reliable route to
alloy SFW prediction, and we discuss the $\gamma$-surface features,
particularly along the shear path sampled during stacking fault
energy calculation, that future interatomic potentials must capture
accurately to enable robust, efficient MD-PFDD modelling of stacking
faults in alloys.

\end{abstract}

\end{frontmatter}


\section{Introduction}

Stacking fault energy (SFE) is an important parameter that is 
key to understanding mechanical behaviour and microstructural 
evolution in metallic systems -- specifically with fcc structure -- affecting 
the strength and ductility of these materials through dislocation-mediated 
phenomena such as: cross-slip, climb, twinning, solute segregation 
(e.g. hydrogen) to the stacking fault plane, which in turn, 
alters the mobility of its bounding dislocations (see, for example, 
~\cite{Rohatgi2001,Youssef2011,Sandlobes2012,Pierce2015,Mahato2015,Yin2016}).  
Experimentally, SFE of a material has, for long, been inferred 
by measuring the equilibrium stacking fault width (SFW), 
which is set by the balance between the elastic energy of the partial 
dislocations and stacking fault energy of the
system~\cite{Whelan1957,Thornton1958,Cockayne1971}. Hence, SFW itself is  
a key parameter governing dislocation dissociation and deformation mode 
selection, particularly in face-centred cubic (fcc) metals and alloys, 
where perfect dislocations spontaneously dissociate into 
Shockley partials bounding a stacking fault. 
Solute segregation to such faults is of particular practical
importance: hydrogen segregation to stacking 
faults~\cite{Kim1985,Pan1986,Shi2024,Yu2025,Xie2021}, for instance, is
implicated in hydrogen embrittlement, one of the most damaging and
difficult-to-predict failure modes in many alloys of importance, making an
accurate, composition-sensitive description of SFW a prerequisite for
predictive modelling of such phenomena.

Molecular dynamics (MD) simulations are well established as capable
of predicting equilibrium SFW in metallic systems with good accuracy,
and have been used extensively to study dislocation dissociation in a
wide range of
metals~\cite{Hakkinen1990,Aslanides1998,Srinivasan2005,Henager2005,Tsuzuki2009} 
and, more recently, alloys~\cite{Yi2017,Rao2017,Shih2021}. 
However, MD is not well suited to phenomena
that require a continuously varying solute concentration field
coupled to the dislocation microstructure, such as solute segregation
to a stacking fault, since this requires prohibitively large simulation 
cells for accommodating dislocations and solutes and long
simulation times to allow diffusive solute redistribution. Phase field
dislocation dynamics (PFDD) offers a more powerful alternative for
such problems: by incorporating the generalized stacking fault energy 
(GSFE) or the $\gamma$-surface calculated from atomistic simulations into the
total free energy functional, PFDD captures the dissociation of
perfect dislocations into partials and the resulting stacking fault,
while remaining naturally compatible with a coupled composition
field~\cite{Shen2003,Shen2004,Hunter2011}. The incorporation of 
$\gamma$-surfaces as a free energy term in PFDD is quite straightforward 
once it is constructed by fitting the stacking fault energies calculated 
at certain characteristic points on the faulting 
planes using DFT or MD simulation~\cite{Schoeck2001,Shen2004,Hunter2011a}. 
The  DFT/MD informed PFDD simulation method is powerful in that it 
enables the simulation of 
dissociation of dislocations into partials and the creation 
of stacking fault between them. Hence, using a concentration 
field that couples with the dislocations, we can 
capture effects such as drag on the glide of partial 
dislocations due to solutes segregating to the stacking faults.

Since Shen and Wang first
incorporated the $\gamma$-surface into a PFDD
framework~\cite{Shen2003}, this approach has been steadily evolving 
and being used to study a number of interesting phenomena, such as, the
dependence of stacking fault width on distance from a grain boundary
ledge~\cite{Hunter2014}, an alternate twinning mechanism in
nanocrystalline fcc metals~\cite{Hunter2014}, the structure and
energy of twist grain boundaries in aluminium~\cite{Shen2014}, and a
twinnability map relating deformation twinning to grain size and
stacking fault energy~\cite{Hunter2015}, in addition to its broader
use in modelling dislocation interactions with precipitates,
inclusions, and grain boundaries~\cite{Varma2023,Wang2010,Beyerlein2016}. 
More recently, PFDD has been extended to incorporate cross-slip of extended 
dislocations~\cite{Wise2025}, incorporate thermally activated dislocation 
motion by combining it with a Monte-Carlo model~\cite{Peng2023}, and 
incorporate dislocation motion in complex concentrated alloys~\cite{Zheng2023} 
and heterogeneous nanostructured materials~\cite{Xu2022}. 

A notable feature in these studies which incorporate $\gamma$-
surface as a term in the free energy of the PFDD model is that the
$\gamma$-surface calculated using density functional
theory (DFT) was predominantly used, and exclusively for pure metals. 
In addition, no study till date, to the best of our knowledge, has extended 
this DFT-informed PFDD approach to an alloy system, despite alloying being 
the precise motivating factor where the use of PFDD over standalone MD -- 
composition-dependent stacking fault energetics and solute segregation to the 
fault -- becomes most relevant. The addition of a solute is well known 
to alter both the magnitude of the stacking fault energy at the characteristic 
points, affecting the $\gamma$-surface, and, correspondingly, the equilibrium 
SFW; yet how well this compositional dependence can be captured by 
atomistically informed PFDD, and by which class of atomistic method, remains 
unaddressed.

Extending this approach to alloys raises questions about which
atomistic method is best suited to provide the $\gamma$-surface.
The DFT based stacking fault calculations, although feasible, 
are compute and time intensive. Classical interatomic potentials 
for binary alloys -- including embedded-atom method (EAM) potentials, 
such as the potentials developed by Liu et al~\cite{Liu1999}, and, Zhou, 
Ward and their respective co-workers ~\cite{Zhou2004,Ward2012} in the case 
of Cu-Al alloys, which are generated by combining single-element potentials 
via efficient alloying schemes -- are typically parameterized against bulk 
equilibrium properties such as lattice parameters, cohesive energies, and 
elastic constants, with the cross-species interaction terms in some 
combination schemes constrained primarily by the properties of ordered 
intermetallic compounds rather than by the defect energetics of the 
disordered solid solution. More recently, machine-learned interatomic 
potentials (MLIPs) such as Gaussian Approximation Potentials (GAP) have 
offered the prospect of DFT-like accuracy at MD-like computational
cost~\cite{parrinello,gap,soap,mishin,mace}; however, the accuracy of
a kernel-based MLIP in any particular region of configuration space
depends entirely on how densely and appropriately that region was
represented in its training database, which is not guaranteed to
include the specific rigid-shift shear path sampled during
$\gamma$-surface calculation. Whether either class of potential --
classical or machine-learned -- can reliably reproduce a DFT-quality
$\gamma$-surface for an alloy system, to a standard sufficient for
downstream PFDD use, is therefore, to be tested.

In this study, we address this question for the Cu-Al system, chosen
as a model fcc alloy because its stacking fault width is known
experimentally to increase with increasing Al
content~\cite{cain1971,Stobbs1971}. We calculate the $\gamma$-surfaces, 
the lattice paramaeters, and elastic moduli as a function of Al composition 
(pure copper to 10 at.\%) using (a) DFT and (b) MD with
two distinct interatomic potentials -- the classical
embedded-atom potential based on the work of Zhou et al~\cite{Zhou2004} 
and Ward et al~\cite{Ward2012} and a machine-learned
GAP potential~\cite{gap}. Each $\gamma$-surface
is incorporated as an energy term into a PFDD model to predict the
equilibrium SFW of edge and screw dislocations, and the results are
cross-validated against standalone MD dissociation measurements, 
isotropic analytical calculations and available experimental data. 
We benchmark the three methods by reproducing bulk lattice and elastic 
properties, and then compare the results of PFDD using
MD and DFT-derived $\gamma$-surfaces to explore the possibility of 
using MD-PFDD instead of DFT-PFDD. The
remainder of this paper is organized as follows: Section~\ref{method} 
details the DFT, MD, and PFDD methodologies used; Section~\ref{result} 
presents thecalculated lattice parameters, elastic moduli, and $\gamma$-surfaces 
from each method, followed by the resulting PFDD- and MD-predicted stacking 
fault widths; and Section~\ref{conclusion}  discusses the implications of 
these results for the development of interatomic potentials suited
to multiscale stacking fault modelling in alloys. 

\section{Computational Methods} {\label{method}}
\subsection{Density Functional Theory}

The first-principles calculations based on density functional theory (DFT)
to optimize the FCC phase structures and obtain the stacking fault
energies of Cu and Cu-rich Al alloys were performed using
the \textit{Vienna Ab-initio Simulation Package}
(VASP) \cite{kresse96a,kresse96b}.
Al solute atoms were randomly substituted at the Cu sites in the simulation 
supercell to get the Cu-Al alloy composition of interest.
We considered 1.67, 4.17, 8.33 and 10.0~at.~\% of Al in Cu,
hereafter, referred to as the Cu2Al, Cu4Al, Cu8Al and Cu10Al, respectively.
In order to accommodate these Al solute percentages,
a supercell of $3\times3\times3$ FCC unit cells was used,
and the bulk structure corresponding to one composition of the
Cu2Al alloy is shown in Fig.~S1 (Supplementary data).
The projector-augmented wave (PAW) psuedo-potentials \cite{blochl94,kresse99}
with the Perdew, Burke, and Ernzerhof (PBE) \cite{perdew96} parametrized
generalized gradient approximation (GGA) were used for the
exchange-correlation functional.
The psuedo-potential with the valence electronic configuration of
Cu: $3d^{10}$ $4s^{1}$ and of Al: $3s^2$ $3p^1$ was used \cite{note1}.
Lattice optimization for each Cu-Al composition was performed to obtain
its equilibrium lattice parameter.
Brillouin zone (BZ) integration was performed with $\Gamma$-centered
Monkhorst-Pack \cite{monkhorst76} mesh based on the Methfessel–Paxton smearing
method with a smearing width of 0.2 eV \cite{methfessel89}.
The plane-wave basis energy cutoff of 600 eV and a $6\times 6\times 6$
k-mesh was used for lattice parameter optimizations.
The total energy was converged to 1 $\mu$eV and the forces on all
atoms to less than 1 meV/\AA.

\subsubsection{Calculation of elastic constants }

The equilibrium structures of pure Cu and Cu-Al alloys were deformed with
a small positive and negative strain ($+$0.02 to $-$0.02) along
the $x$, $y$, $z$, $xy$, $xz$, $yz$ directions to obtain its energy
as function of applied strain.
The energy-strain relation in each direction was used to obtain the
three elastic constants ($C_{11}$, $C_{12}$, and $C_{44}$) as described
in Ravindran {\it et. al.} \cite{ravindran98}.
Each elastic constant was calculated by fitting a polynomial to the total
internal energy as a function of strain parameter $\delta$.
The elastic constant is identified as the second order coefficient in
the energy-strain relation given by the Taylor series expansion in
terms of small deformation $\delta$ in a regime of Hooke’s law:
\begin{equation}
E(V,\delta)=E(V_0,0)+V_0\Bigl(\sum_i \tau_i\epsilon\delta_i
+\frac{1}{2}\sum_{ij}C_{ij}\delta_i \delta_j \epsilon\Bigr)+O(\delta^3)
\label{strain}
\end{equation}
Here, $E(V,\delta)$ and $E(V_0,0)$ is the total internal energy of
strained and unstrained crystal, respectively. $V_0$ is the volume of
unstrained crystal and $C_{ij}$ is the elastic constant.
$\tau_i$ is the stress tensor element that is to be obtained by fitting
a polynomial to the energy-strain curve.
An example of our calculated energy-strain curves with the fitted
function are shown in Fig.~S2 (Supplementary data).

\subsubsection{Calculation of $\gamma$-surface}
\label{sec:gamma-surface}

The composition dependent optimized bulk lattice parameter was used to
generate a crystal slab structure with 10 $\lbrace$111$\rbrace$ layers
having $ABCABC...$ stacking sequence in a simulation cell~\cite{note2}.
The layers were ensured to be periodic in the $x$ and $y$ directions
and a 15 \AA~ vacuum in the $z$ direction was added to avoid interactions
in the non-periodic dimension between periodic images.
Dislocation in the slip plane was modeled by displacing the
upper 5 layers in the $\langle 110 \rangle$ 
or the $\langle 112 \rangle$ directions.
The structure of the slab without and with the stacking fault for Cu is
shown in Fig.~S3. Fig.~S4(a)-(d) show the structures of the perfect
crystal slab of Cu-Al alloys at the different Al concentrations.

The generalized stacking fault energy (GSFE) is defined as the total
energy difference required per unit area to generate a single layer
stacking fault in the $\lbrace 111\rbrace$ slip plane by shearing one elastic
half space of the crystal relative to the other along the $\langle112\rangle$
slip direction \cite{vitek68}. It is given as:
\begin{equation}
\gamma_{GSFE}^{(f)} = \frac{E^{(f)}-E^{(p)}}{\Omega}
\end{equation}
where, $E^{(f)}$ is the internal energy of the supercell  with a
stacking fault equal to the magnitude of the fault vector ($f$)
in the glide plane. $E^{(p)}$ is the internal energy of the perfect crystal
structure, and, $\Omega$ is the surface area of the  perfect crystal
structure. To obtain the stacking fault energies at the special points
on the $\gamma$ surface denoted by $A$, $G_1$, $G_2$, $G_3$, $G$, $T_1$,
$T$~\cite{Shen2004}, we varied the stacking fault vector ($f$) 
from $0.0\textbf{b}$ to $1.0\textbf{b}$ with a step of $0.25\textbf{b}$
along the $\langle$112$\rangle$ direction ($G_{1}$ $(0.25\textbf{b})$, $G_{2}$
$(0.5\textbf{b})$, $G_{3}$ $(0.75\textbf{b})$ and $G$ $(1.0\textbf{b})$); and,
from $0.0\textbf{b}$ to $0.5\textbf{b}$ with a step size of $0.125\textbf{b}$
for $T_1$ $(0.125\textbf{b})$ and $T$ $(0.25\textbf{b})$ points along the
$\langle$110$\rangle$ direction, where $\textbf{b}$ is the
magnitude of Burgers vector.

A simulation cell of $2\times3$ unit cells in the x, y dimension with
10 layers containing 120 atoms for Cu, Cu2Al and Cu4Al, and
$3\times3$ unit cells with 10 layers containing 180 atoms for
Cu8Al and Cu10Al were used for stacking fault energy calculations.
Al atoms were distributed randomly in the simulation
cell in the different layers. The random assignment of Al atoms
is shown in Table~S1.
In the calculations of GSFEs,
an optimized $8\times6\times1$ k-mesh was used, due to its better
agreement with the experimental intrinsic stacking fault energy.
In these calculations, all atoms in the two top and bottom surface layers 
were held fixed in all directions to their bulk configurations.
In the remaining layers, atoms were kept fixed along the $x$
($\langle$112$\rangle$) and $y$ ($\langle$110$\rangle$) directions
and the atomic positions were relaxed only along the $z$
($\langle$111$\rangle$) direction by minimization of
Hellmann-Feynman forces per atom. The calculated values of
these GSFEs were used to obtain the coefficients that
represent the $\gamma$ surface of an alloy composition.

\subsection{Molecular Dynamics}
The Large-scale Atomic/Molecular Massively Parallel Simulator 
(LAMMPS)~\cite{plimpton1995} is used to carry out molecular dynamics 
simulation (MD). The ZW embedded atom method 
potential~\cite{Zhou2004,Ward2012} and the machine learning based potential 
named the Gaussian approximation potential (GAP)~\cite{Antillon2025} is 
used to calculate the lattice parameters, elastic constants and GSFE values 
at the characteristic points, as described in sections~\ref{elasML}  and 
\ref{gsfeML}. The SFW is also calculated using standalone MD simulations 
based on the ZW potential as described in section
~\ref{sec:sfw-standalone-md}.

In addition, we also calculated the elastic constants and 
stacking fault energies using the MACE-MP-0 foundation model, 
trained on an open-source DFT dataset (MPTrj) of material structures and 
physical properties~\cite{mace}. The size of the simulation boxes in the 
simulations based on MACE potential is small due to our GPU hardware 
limitations. However, the MACE potential is not used for the multiscale 
PFDD investigation of SFW due to the extreme deviation of the GSFE values from the corresponding DFT-calculated values. The calculated 
values are reported in the Supplementary Information.

\begin{figure}[h]
    \centering
    \includegraphics[scale=0.5]{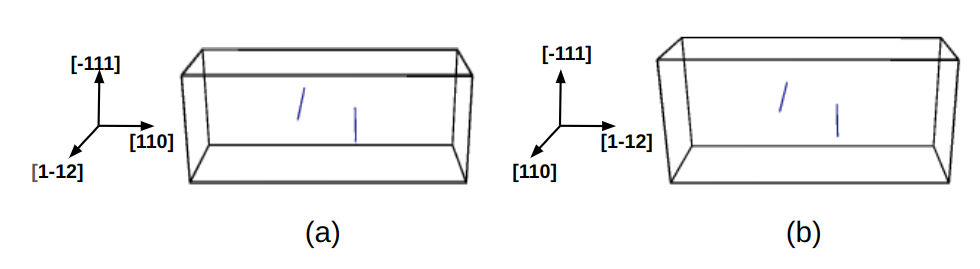}
    \caption{The initial perfect dislocation dipole configuration (atoms are removed for a visualisation of dislocations) (a) edge (b) screw. }
    \label{initial_configuration}
\end{figure}

The simulation cell with crystal orientations  $\left[110\right]$, $\left[\overline{1}11\right]$, and $\left[1\overline{1}2\right]$ had dimensions $120 \times 14 \times 20$ unit cells for edge dislocation, as shown in Fig.~\ref{initial_configuration}(a). The screw dislocation with orientations $\left[1\overline{1}2\right]$, $\left[\overline{1}11\right]$, and  $\left[110\right]$ had dimensions $70 \times 22 \times 40$ unit cells as shown in Fig.~\ref{initial_configuration}(b). The periodic boundary conditions is used in all dimensions for both the configurations. In this study, the dislocation dipole configuration is considered for calculating the stacking width because the sum of the Burgers vector is zero; therefore, the long-range elastic interaction cancels out and periodic boundary condition is maintained. The crystallographic details of the dislocation are as follows: the Burgers vector of the dislocation is $\frac{a}{2}\left[110\right]$, $\left[\overline{1}11\right]$ is normal to the glide plane and the line direction of the dislocation is $\left[1\overline{1}2\right]$ as shown in Fig.~\ref{initial_configuration}. The simulation cell contains pure Cu atoms and Cu-Al alloys with 2, 4, 8, and 10 atomic percent Al, with Al atoms randomly distributed in pure Cu.

The initial configuration is minimized using the conjugate gradient method by setting the energy and force tolerance values to $10^{-25}$ and the maximum number of iterations to $10^5$. Then, the microstructure after minimization is equilibrated under the microcanonical ensemble (NVE) for 300 ps, with a snapshot saved every 1 ps during equilibration.

\subsubsection{Calculation of the lattice parameters and moduli}{\label{elasML}}

The lattice parameter and elastic moduli for Cu and Cu-alloys at 0 K are calculate using a simulation box with orientation along x, y and z is $\left[110\right]$, $\left[\overline{1}11\right]$, and $\left[1\overline{1}2\right]$. The simulation box is minimized using conjugate gradient method having energy tolerance and force tolerance $1\times 10^{-25}$. The length of simulation box $L_x$, $L_y$ and $L_z$ of minimized configuration is taken to obtain lattice parameter from each dimension and average is taken to obtain equilibrium lattice parameter. To calculate the elastic constants, Hooke's law is used:
\begin{equation}
    \sigma_{ij} = C_{ijkl}\varepsilon_{kl}
\end{equation}
The stress and strain is required to calculate the elastic constant using Hooke's law. The simulation box is oriented along the $x,y,z$-axes
in the $\left[100\right]$, $\left[010\right]$, and $\left[001\right]$ directions, respectively, and then minimized as described. The simulation box
is deformed in both positive and negative directions along
the normal (x, y, and z) and shear (xy, xz, and yz) directions.
The box is displaced to induce a strain of 0.2\%, and the resulting stress
tensor is obtained from the deformation. The final values for the elastic
constants are obtained by taking the average of the constants derived
from the positive and negative loading.

\subsubsection{Calculation of the $\gamma$-surface} {\label{gsfeML}}

The generalized stacking fault energy (GSFE) is the energy required to displace atoms between two adjacent planes in a specific direction. The simulation box of copper and copper alloys are created and the box is split into two regions top and bottom along $\left[\overline{1}11\right]$ direction. The upper half is displaced along $\left[1\overline{1} 2\right]$ by increments of 0.1 \AA, and after each displacement, the simulation box is relaxed, and the potential energy is computed. The GSFE is calculated by taking the energy difference between the final state (after displacement) and the initial state (before displacement), divided by the area of the plane along a shear direction. 
\subsubsection{Calculation of stacking fault width} {\label{sec:sfw-standalone-md}}

The perfect dislocation with the Burgers vector $\frac{a}{2}\left[110\right]$ dissociates into two Shockley partial dislocations to minimize potential energy. This dissociation occurs to minimize elastic strain energy at the expense of the energy required to create a faulted plane, in accordance with Frank's criterion. The stacking fault width, representing the equilibrium separation between the two Shockley partial dislocations, is calculated for
Cu, Cu2Al, Cu4Al, Cu8Al, and Cu10Al by measuring the distance between the two Shockley partial dislocations in the equilibrated microstructure. The width is determined by discretizing the dislocation line into points, measuring the distance between these points, and taking an average. 
\subsection{Phase field dislocation dynamics}

In the phase field dislocation dynamics (PFDD) model, 
the order parameter ($\eta_{ss}$) defines the slipped 
region corresponding to a particular Burgers vector in a slip plane.
Dislocations are denoted by diffuse interfaces that differentiate between slipped regions and unslipped regions in the slip plane. 
The plastic strain associated with a system consisting of dislocations
in $n$ slip systems can be written in terms of the order parameter as:

\begin{equation}
    \epsilon^{pl}_{ij}(\mathbf{x}) = \sum_{ss=1}^{n}\left[\frac{b_{ss,i} n_{ss,j} + b_{ss,j} n_{ss,i}}{2}\right]\eta_{ss}(\mathbf{x})
\label{eigenstrain}
\end{equation}

where $b_{ss,i}$, $n_{ss,i}$ are the 
components of the Burgers vector and the slip plane 
normal respectively, and, $\eta_{ss}$ is the order parameter
corresponding to the slip system $ss$.

The total energy of the system containing dislocations is given by:
\begin{equation}
    E^{tot} = E^{el} + E^{gsf}
\end{equation}
where $E^{el}$ is the elastic energy and $E^{gsf}$ is the
generalized stacking fault energy. 

The elastic energy is calculated by the following volume integral:
\begin{equation}
    E^{el} = \frac{1}{2}\int C_{ijkl}\epsilon^{el}_{ij}(x)\epsilon^{el}_{kl}(x)\; dV
\end{equation}
where $C_{ijkl}$ is the modulus of elasticity, and, 
$\epsilon^{el}_{ij}(x)$ is the elastic strain field 
due to the dislocation at a given point, obtained as:
\begin{equation}
\epsilon^{el}_{ij}(x) = u_{i,j} - 
\epsilon^{pl}_{ij}(x)   
\end{equation}
where $u_i$ is the displacement field due to the dislocations. $u_i$ is obtained
by solving the Eshelby problem~\cite{Mura1987} with the eigenstrain equal to
the plastic strain given in eq. \ref{eigenstrain}.

In addition to the elastic energy, which takes care of dislocation-dislocation
interactions, energy is also required to move the dislocations through
a crystal lattice. This energy component is called the generalized stacking
fault energy (GSFE), and is defined as \cite{Shen2003}:
\begin{equation}
E^{gsf} = \sum_{ss=1}^n \int_{ss} \mathcal{F}_{ss}[\eta_{ss}(x)]\; d^2x
\end{equation}
where $\mathcal{F}_{ss}(\eta_{ss}(x))$ is a periodic 
potential function that represents the energy 
landscape of the slip system $ss$, in which the 
dislocation has to glide.

We consider a face-centered cubic system, where slip
occurs mainly in close-packed planes $\lbrace 111 
\rbrace$ along the close-packed directions $\langle 
110 \rangle$. Specifically, we define three
order parameters ($\eta_{i},\;i=1,2,3$) corresponding
to three Burgers vectors $b_1  = 
\frac{1}{2}[0\bar{1}1],\;b_2 = 
\frac{1}{2}[10\bar{1}],\;\textrm{and}\;b_3 = 
\frac{1}{2}[\bar{1}10]$ in the slip plane $(111)$. 
Hence, displacement along any direction in the slip plane can be represented
by a linear combination of the three Burgers vectors~\cite{Shen2003}:
\begin{equation}
\Delta_s = \left[ \eta_1(x)\hat{b}_1 + \eta_2(x)\hat{b}_2 + \eta_3(x)\hat{b}_3\right] b
\label{eq:cumul-disp}
\end{equation}
where $b$ is the magnitude of the Burgers vector in the $(111)$ plane and
$\hat{b}_i$ is the unit vector along each slip direction. This is important
because in fcc systems, it is energetically favorable for perfect dislocations
to dissociate into Schockley partials, creating a stacking fault between the 
partials. This is captured by the $\frac{d \Delta_s}{dx}$ profile, in which, 
the two peaks indicate the positions of the Shockley partial dislocations. The 
dissociation of dislocations and the creation of stacking fault can be captured 
in the model by using a periodic potential function that represents the 
$\gamma$-surface.

The following form of the potential function, written in terms of sines
and cosines, given by \cite{Schoeck2001},
has been used to represent the $\gamma$-surface.
\begin{equation}
\begin{split}
\mathcal{F}(\eta_1,\eta_2,\eta_3) = c_0 &+
c_1\left[ cos2\pi(\eta_1-\eta_2) +                                cos2\pi(\eta_2-\eta_3) +                                cos2\pi(\eta_3-\eta_1) \right] \\&+ 
c_2\left[ cos2\pi(2\eta_1-\eta_2-\eta_3) +                        cos2\pi(2\eta_2-\eta_3-\eta_1) \right.\\&\left.+\; 
          cos2\pi(2\eta_3-\eta_1-\eta_2) \right] +
c_3\left[ cos4\pi(\eta_1-\eta_2) \right.\\&\left.+\;                cos4\pi(\eta_2-\eta_3) +                                cos4\pi(\eta_3-\eta_1) \right] \\&+
c_4\left[ cos4\pi(3\eta_1-\eta_2-2\eta_3) +                       cos4\pi(3\eta_1-2\eta_2-\eta_3) \right.\\&\left.+\;                                      cos4\pi(3\eta_2-\eta_3-2\eta_1) +
          cos4\pi(3\eta_2-2\eta_3-\eta_1) \right.\\&\left.+\;
          cos4\pi(3\eta_3-\eta_1-2\eta_2) +
          cos4\pi(3\eta_3-2\eta_1-\eta_2) \right] \\&+ 
a_1\left[ sin2\pi(\eta_1-\eta_2) +                                sin2\pi(\eta_2-\eta_3) \right.\\&\left.+\;                sin2\pi(\eta_3-\eta_1) \right] + 
a_3\left[ sin4\pi(\eta_1-\eta_2) \right.\\&\left.+\;                sin4\pi(\eta_2-\eta_3) +                                sin4\pi(\eta_3-\eta_1) \right] 
\end{split}
\label{gammaSEquation}
\end{equation}

The coefficients $c_i$ and $a_i$ are obtained by fitting
to the $\gamma$-surface obtained from first principle simulations (explained in section \ref{sec:gamma-surface}).
The coefficients can be written in terms of the seven points as~\cite{Schoeck2001}:

\begin{equation}
\begin{split}
    c_0 &= 0.823[4G-6G_1 + 6G_2 -7.392G_3 + 0.804T + 0.804T_1] \\
    c_1 &=0.274[ -8G + 12G_1-12G_2 + 14.785G_3 -1.608T + 0.215T_1]  \\
    c_2 &= 0.091[ 23.072G-29.138G_1+32.785G_2-42.215G_3+2.569T-2.412T_1] \\
    c_3 &= 0.137[-8G + 12G_1-12G_2 + 14.785G_3 + 0.215T -1.608T_1]  \\
    c_4 &= 0.023[1.856G-13.723G_1+6.431G_2-4.277G_3-0.962T+3.531T_1]  \\
    a_1 &= 0.137[-32G+48G_1-48G_2+62.785G_3-4.608T-2.785T_1]  \\
    a_3 &= 0.046 [17.072G-19.292G_1+31.923G_2-34.708G_3+3.341T-8.354T_1] 
\end{split}    
\end{equation}

The three order parameters $\eta_i\;i=1,2,3$ are evolved using the time
dependent Ginzburg Landau equation:
\begin{equation}
\frac{\partial \eta_i}{\partial t} = -L \frac{\delta E^{tot}}{\delta \eta_i}
\end{equation}

\section{Results and Discussion}{\label{result}}

\subsection{Lattice parameter and elastic constants}

\begin{figure}[h]
    \centering
    \includegraphics[width=\linewidth]{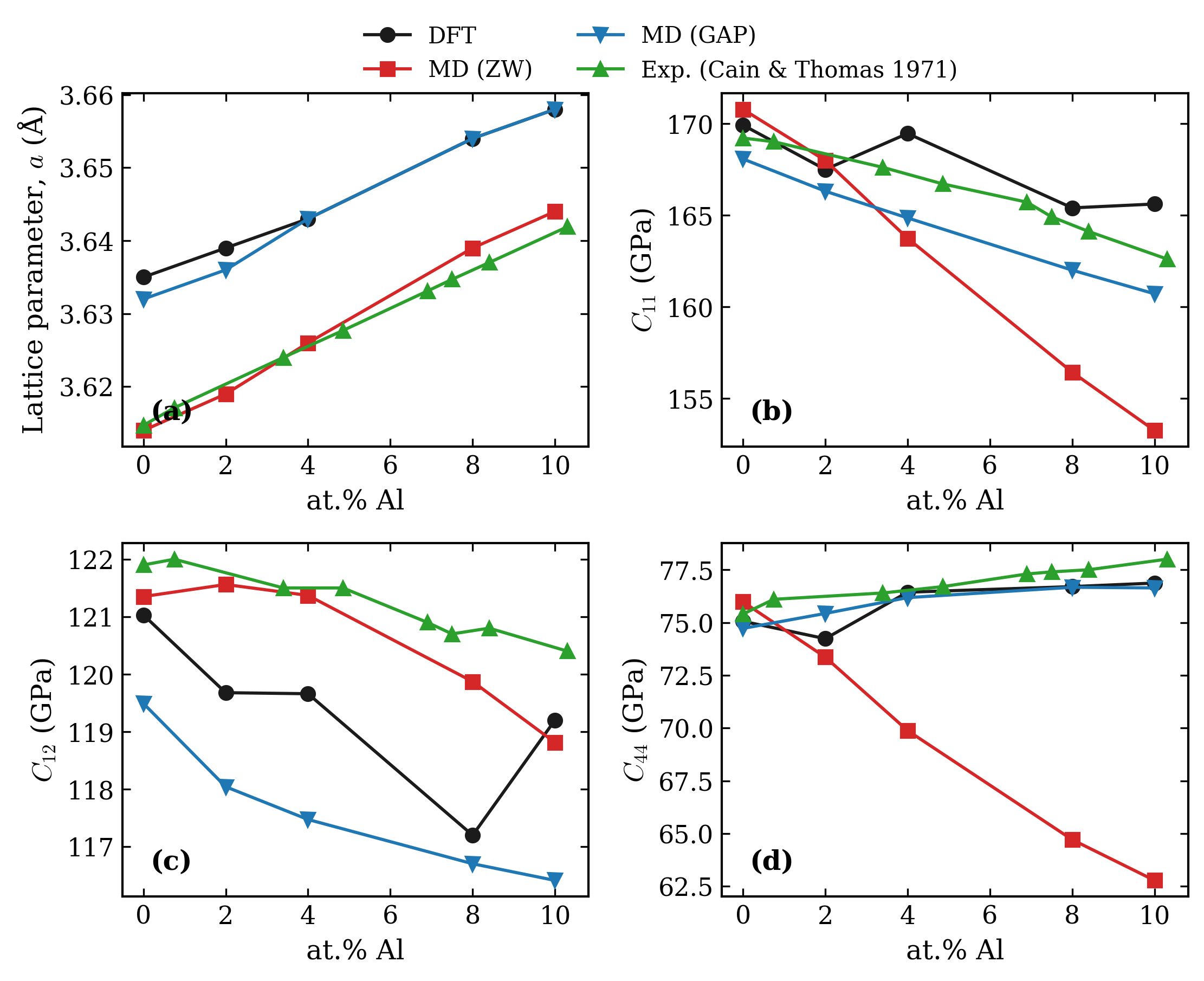}
    \caption{Lattice parameter and elastic constants calculated using DFT and MD compared with experimental measurements from~\cite{cain1971}. The lines connecting the points are meant to be a guide to the eye.}
    \label{fig:ModuliLatticeParameter}
\end{figure}

Fig.~\ref{fig:ModuliLatticeParameter} shows the lattice 
parameters and the moduli components $C_{11},\;C_{12}\;
\textrm{and}\;C_{44}$ calculated using DFT and MD simulations 
(using ZW and GAP potentials), compared with 
experimental observations of the same~\cite{cain1971}. There 
is a resaonable agreement between the properties calculated 
from the simulations and the experimental measurement in that 
the trends are captured well (except, of course, the case of $C_{44}$). 

The difference in the lattice parameters calculated by DFT and 
MD (GAP), as compared to the experimental measurement, is 
expected. The PBE exchange correlation functional is
known to overestimate the lattice parameter. As the GAP 
potential is also on a set of reference DFT configurations, 
the overestimation in the lattice parameter calculation is 
expected in this case as well. On the contrary, there is an 
excellent agreement between the experimental values and the MD 
(ZW) case. ZW potential is a classical EAM potential 
whose parameters are fit directly against experimental values 
of properties. 

The moduli components also largely reproduce the same trend as 
in the experiments~\cite{cain1971}, with DFT simulations and 
MD (GAP) simulations showing very close agreement. 
The fluctuations in the moduli components from DFT simulation 
for the alloy cases could have their origins in the configurational 
sampling in the finite-sized simulation cells. 
Although the decreasing trend is correctly captured 
for the $C_{11}$ and $C_{12}$ cases by the ZW and GAP 
potentials, there is a wider deviation from the 
experiments, specifically for MD (ZW)-calculated $C_{11}$. 
Further, the $C_{44}$ trend predicted by the MD (ZW) 
contradict the experimental observations; this anomaly is noted 
in previous studies as well~\cite{Kamalakshi2022,Moment1972}. 
The numerical values 
of the lattice parameters and elastic moduli ($C_{11}$, 
$C_{12}$, and $C_{44}$) obtained from DFT and MD 
simulations for pure Cu and Cu--Al alloys are provided in 
Table~S2 in the Supplementary Information. We also note that the 
agreement of the DFT-calculated elastic moduli with the experiment may be 
improved by employing higher-rung exchange-correlation functionals, 
such as RSCAN meta-GGA functionals, but of course, with a 
higher computational overhead.

\subsection{Equilibrium stacking fault widths from standalone MD}

\begin{figure}[h]
    \centering
    \includegraphics[width=15cm]{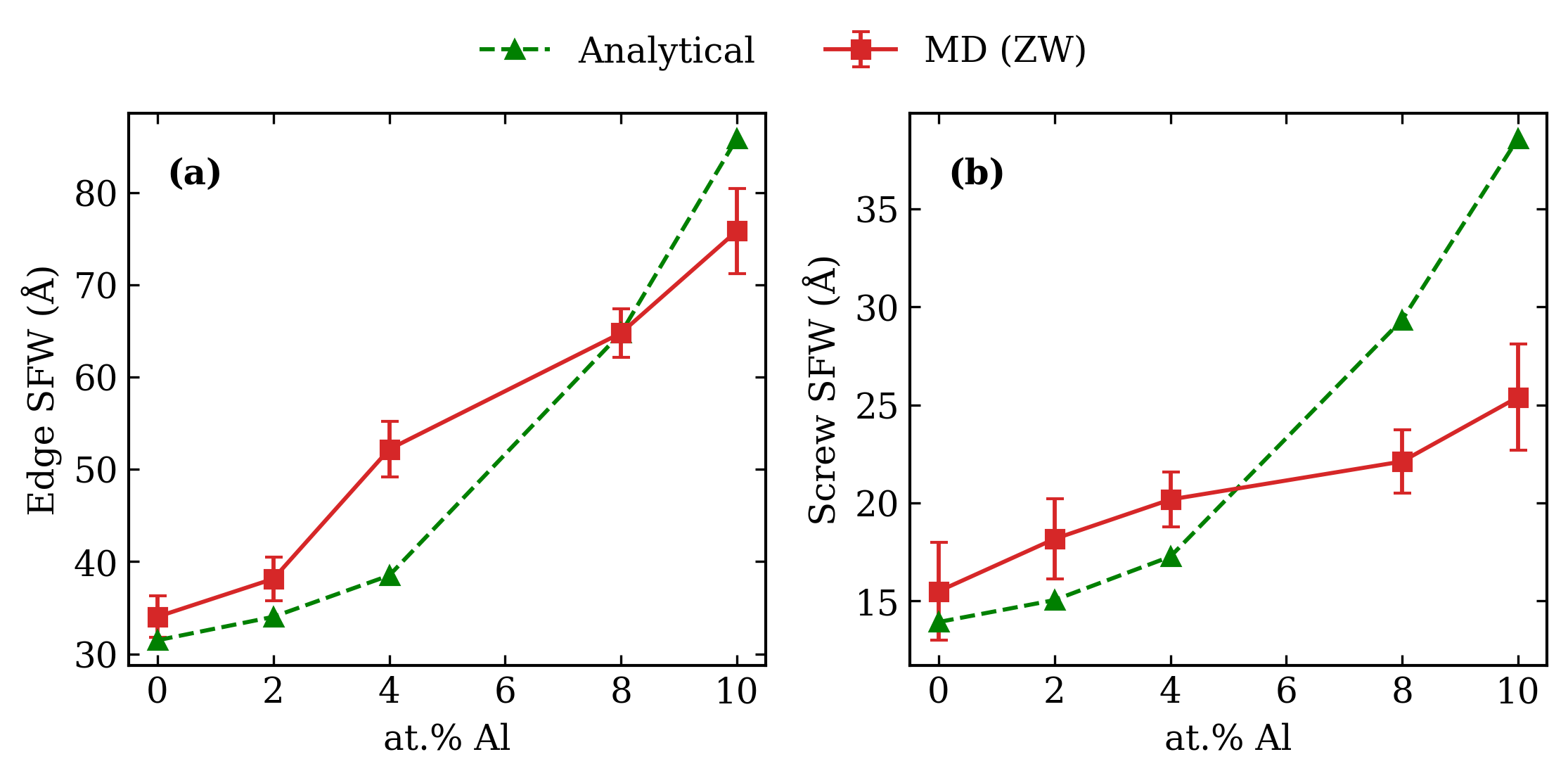}
    \caption{Comparison of the intrinsic stacking fault width calculated from standalone MD with Eq.~(\ref{sfw_eq}). The lines are meant as a guide to the eye.}
    \label{fig:sfw-md}
\end{figure}

Next, we calculated the intrinsic stacking fault width for 
edge and screw type (straight) dislocations using standalone 
MD simulations using the ZW potential. The average 
distance between the dissociated partials is calculated for 
different Al concentrations are shown in Fig.~\ref{fig:sfw-md} 
and compared with the analytical calculations of the 
stacking fault width according to the relation:
\begin{equation}
w^{eq}_{SF} = \frac{\mu}{2\pi\gamma_I}\Bigg[ (b_L\cdot \xi_L)(b_T\cdot\xi_T) + \frac{(b_L \times \xi_L)(b_T \times \xi_T)}{(1-\nu)}\Bigg]
\label{sfw_eq}
\end{equation}
where $\mu$ is the shear modulus, $\gamma_I$ is the intrinsic
stacking fault energy (SFE), and $b_T$, $b_L$ and $\xi_L$, 
$\xi_T$ are the Burgers vectors and sense vectors of the 
leading and trailing dislocations respectively. Note that 
this relation assumes that $w^{eq}$ is a function of only the 
$\gamma_I$ and not of all the six characteristic points on
the $\gamma$-surface. The elastic parameters, 
namely shear modulus ($\mu$) and Poisson's ratio ($\nu$) are 
obtained from elastic constants $C_{ij}$ (shear modulus by
Voigt averaging), intrinsic stacking fault energy ($\gamma_I$) and 
magnitude of the Burgers vector ($b$) are obtained using 
MD simulations at the same concentrations. 

For the pure-Cu case, there is an excellent agreement between the SF widths 
calculated from the MD simulations and the analytically 
calculated values. The SFW values for 
the pure Cu case also compare well with the experimentally 
measured values of $38$ $\pm$  $6$~\AA~and $18$ $\pm$ 
$6$~\AA~for edge and screw cases, 
respectively~\cite{Stobbs1971}. In Eq.~(\ref{sfw_eq}), which has been 
generally used to estimate stacking fault widths in systems 
based on the other properties, the 
equilibrium SF width is considered to depend only on the 
intrinsic stacking fault energy and not on the entire $\gamma$-
surface~\cite{Schoeck2001}. In addition, 
the equation assumes isotropy of elastic moduli, which is not 
the case in Cu-Al system that has an increasing anisotropy 
with Al addition. This assumption of elastic isotropy is known to 
overestimate the stacking fault width for screw type dislocations~\cite{Szajewski2018}
In spite of these differences, the stacking fault 
widths are quite close to the analytical values obtained from 
Eq.~(\ref{sfw_eq}) for the pure Cu 
and the dilute alloy concentration of upto Cu-4Al, with the deviations
becoming more apparent only for the larger alloy compositions.

\subsection{$\gamma$-surfaces from DFT versus MD}

The stacking fault energy values at the seven significant 
points calculated using MD simulations (with three different 
interatomic potentials) and DFT simulations are shown in 
Table~\ref{tab:gsfe-points-dft} and detailed comparison
is given in Table~S3 in Supplementary Information.
Points G and G$_2$ represent 
the intrinsic and extrinsic stacking fault energies 
respectively for each alloy. As expected, both intrinsic and 
extrinsic SFEs show a downward trend for increasing Al 
content. There is a noticeable difference between the energy 
values calculated using MD and DFT methods for almost all 
points. In particular, there is a stark difference for values 
calculated at points T and T$_1$. 

\begin{table}[h]
   \centering
 \caption{The energy values (in $mJ/m^2$) at the seven  points labelled 
    on the GSFE surface (see section~\ref{sec:gamma-surface}) and calculated using DFT and MD simulations.}
    \begin{tabular}{lllllllll}
    \hline
    Material & Sim. type & $A$ & $G_1$ & $G_2$ & $G_3$ & $G$ & $T_1$ & $T$ \\
    \hline   
    Cu & DFT & 0 & 82.66 & 156.27 & 108.37 & 38.85 & 288.53 &  513.99 \\ 
     & MD (ZW) & 0 & 83.76 & 178.95 & 112.95 & 40.21 & 344.68  & 710.03 \\
     & MD (GAP) & 0 & 93.33 & 185.76 & 125.53 & 40.26 & 358.88  & 756.00 \\
     \hline
    Cu2Al & DFT & 0 & 83.79 & 153.63 & 98.37 & 23.43 & 284.76 & 499.33 \\ 
     & MD (ZW) & 0 & 79.65 & 170.85 & 106.25 & 35.90 & 334.41 & 696.11 \\
      & MD (GAP) & 0 & 92.49 & 183.31 & 121.90 & 35.87 & 356.03 & 751.01 \\
      \hline
    Cu4Al & DFT & 0 & 83.75 & 152.05 & 93.38 & 13 & 283.82 & 497.36 \\ 
     & MD (ZW) & 0 & 74.05 & 159.38 & 96.63 & 29.94 & {325.43} &{675.44} \\ 
     & MD (GAP) & 0 & 90.61 & 177.85 & 114.10 & 28.22 & 349.49 & 738.32 \\
     \hline
    Cu8Al & DFT & 0 & 81.03 & 146.97 & 88.10 & 9.92 & 275.66 & 482.25 \\ 
     & MD (ZW) & 0 & 66.94 & 143.20 & 79.89 & 16.37 & 304.00 & 649.97 \\
     & MD (GAP) & 0 & 88.27 & 172.07 & 107.08 & 19.45 & 341.31 & 725.11 \\
     \hline
    Cu10Al & DFT & 0 & 79.79 & 144.43 & 86.00 & 8.63 & 273.28 & 478.81 \\
     & MD (ZW) & 0 & 65.59 & 140.12 & 76.23 & 12.03 & 301.04 & 645.69 \\
     & MD (GAP) & 0 & 86.21 & 166.06 & 99.16 & 12.80 & 336.33 & 715.03 \\
    \hline
    \end{tabular}
    \label{tab:gsfe-points-dft}
\end{table}

\begin{figure}[h!]
    \centering
    \includegraphics[width=18cm]{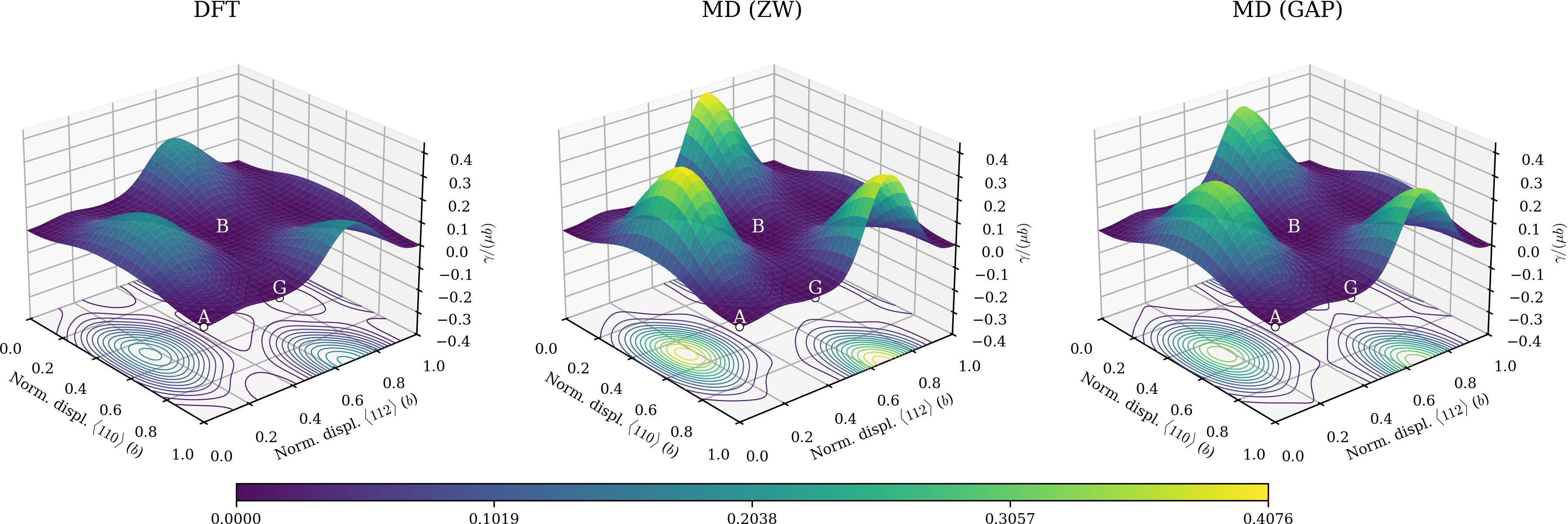}
    \caption{Normalized GSFE surfaces obtained by fitting the Fourier expression to the characteristic points $A,\;G_1,\;G_2,\;G_3,\;G,\;T_1,\;\textrm{and}\;T$ from DFT, MD (ZW) and MD (GAP) calculations for the case of Pure Cu. The minima at A and B and the local minima at G are marked on the surfaces.}
    \label{fig:gsfe-surface}
\end{figure}

\begin{figure}[h!]
    \centering
    \includegraphics[width=0.8\linewidth]{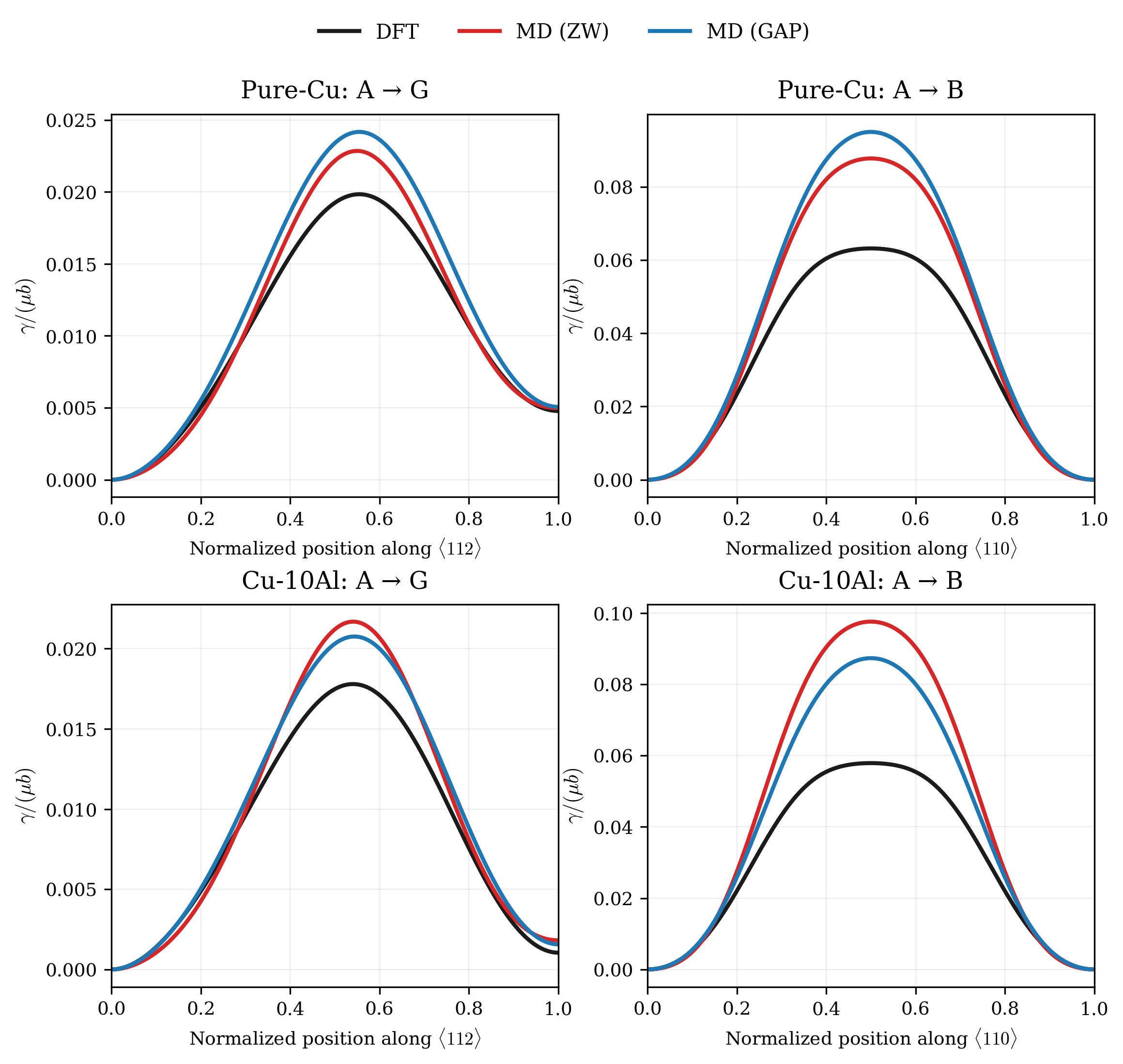}
    \caption{Stacking fault energy profiles plotted along the  Burgers vector $\overrightarrow{AB}$ of the perfect dislocation and the Burgers vector $\overrightarrow{AG}$ of the partial dislocation for Pure-Cu and Cu-10Al cases respectively. Note that the positions are normalized with respect to the magnitude of $\vec{b}$ of the partial dislocation and the full dislocation located along the paths $A\rightarrow G$ and $A\rightarrow B$.}
    \label{fig:gsfe-profile}
\end{figure}

The $\gamma$ surfaces constructed by fitting to the seven significant points 
(labelled in section~\ref{sec:gamma-surface}) calculated 
using DFT and MD simulations are shown in 
Fig.~\ref{fig:gsfe-surface}. The surfaces indicate the 
minima at A and B ($\overrightarrow{AB}$ is the Burgers vector 
of the perfect dislocation) and the local minimum at G ($\overrightarrow{AG}$ and 
$\overrightarrow{GB}$ are the Burgers vectors of the leading and trailing partials). It is obvious that the magnitudes of the peaks are higher in the MD calculations, 
with the highest peaks observed in the MD (ZW) case. For a clearer picture of the saddle 
points and minima along the Burgers vectors, we have also plotted the energies along these 
directions in Fig.~\ref{fig:gsfe-profile}, from which the stark 
difference in the glide energy barriers of both partial and perfect dislocations is 
evident. 
\begin{figure}[h!]
    \centering
    \includegraphics[width=0.99\linewidth]{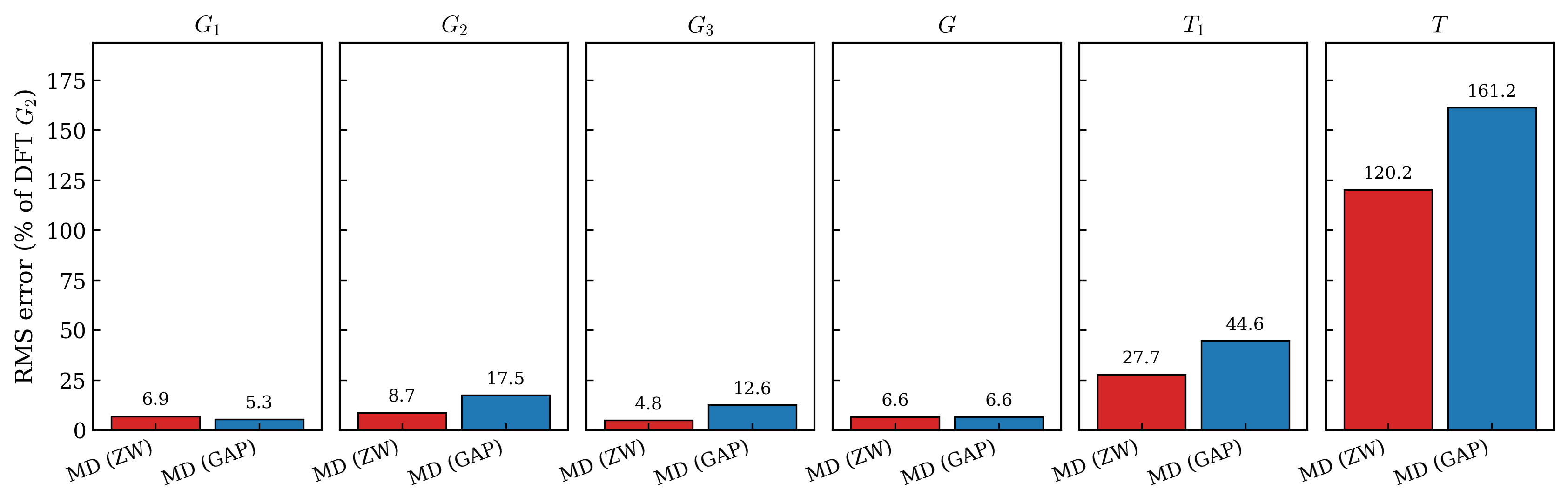}
    \caption{Root mean square deviation of stacking fault energies of all alloy compositions (computed at the characteristic points) by MD relative to DFT. The unstable stacking 
    fault energy computed by DFT case is used as reference for each alloy composition.}
    \label{fig:errorbar}
\end{figure}
Furthermore, to quantify the deviation of the MD-based $\gamma$-surfaces from the DFT-
based $\gamma$-surface, we depict the root mean square deviation (relative to the DFT, as 
a fraction of the unstable stack fault energy ($G_2$) from DFT) for all alloy 
compositions considered using both MD (ZW) and MD (GAP) in Fig.~\ref{fig:errorbar}. Clearly, the GAP potential seems to have higher or comparable errors in contrast with the 
ZW potential. The most significant errors occur at points $T$ and $T_1$. Thus, even though the interatomic potentials used in MD get the stacking fault energy itself correct, the energy landscape, especially with respect to the barrier heights is not accurate. This, as we see in the PFDD simulations, leads to differences in partial separation as well as pinning of dislocations due to the changed 
the energy barrier.

\subsection{Stacking fault widths from the GSFE incorporated PFDD model}

Finally, we incorporated the $\gamma$ surfaces obtained from the DFT and MD 
simulations into the PFDD model to observe dissociation of dislocations and 
stacking fault formation. The following results are from simulations 
performed on a $512\times512$ grid, with a grid spacing equal to the 
magnitude of the Burgers vector in each case. The initial configuration 
considered is that of a perfect dislocation dipole (of edge or screw 
character) with a separation of 256$\vec{b}$, to minimize the interaction 
with the opposite dislocation. 

\begin{figure}[h!]
    \centering
    \includegraphics[width=0.8\linewidth]{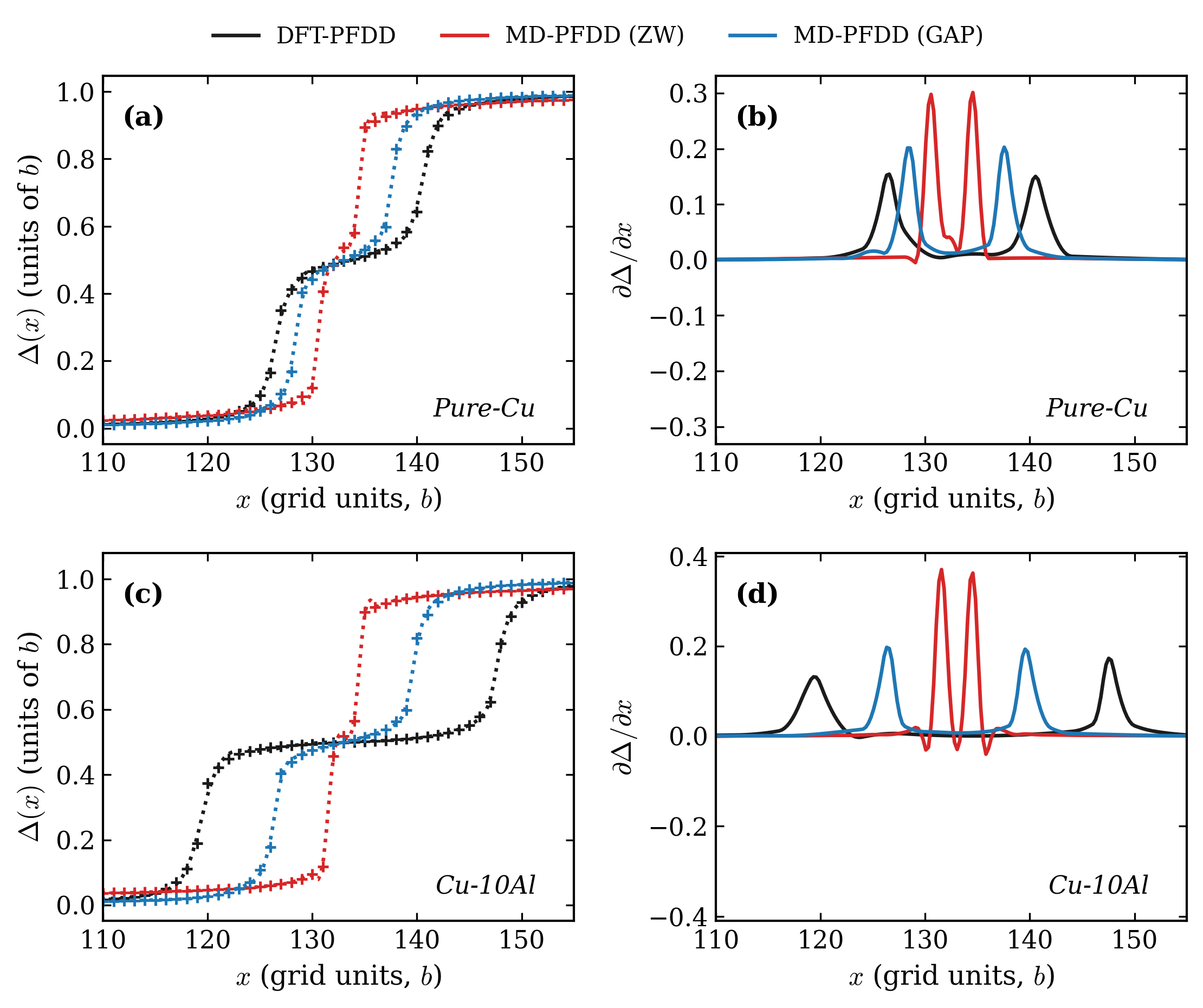}
    \caption{The cumulative displacement profiles Eq.~(\ref{eq:cumul-disp})) from the simulations post equilibration, along with a spline fit and a gradient calculated to locate the partial separations. The first row shows the results for Pure-Cu and the second row shows the Cu-10Al case.}
    \label{fig:pfdd-partial-separation}
\end{figure}

In Fig.~\ref{fig:pfdd-partial-separation}, we show, for pure Cu and 
Cu-10Al, the cumulative displacement profile calculated as per 
Eq.~(\ref{eq:cumul-disp}), from the slip order parameters evolved using 
PFDD simulations supplied with DFT- and MD-based $\gamma$-surfaces. The 
dissociation of the dislocations into partials, evidenced by the two peaks 
in the $\frac{\partial \Delta}{\partial x}$ plots, is observed in all the 
simulation cases. However, the equilibrium stacking fault width is lower 
in MD-based PFDD calculations compared to DFT-PFDD. The height of the peaks 
are also higher in MD-based PFDD due to the high SFE values at the barriers, 
that resist the spreading of the dislocation core.

Before discussing the equilibrium stacking fault width, it needs to be
ensured that the dislocations were not pinned to the grid and were indeed 
mobile. For this, we performed another set of simulations with
inter-dislocation separation reduced to 64$\vec{b}$. This separation was 
chosen due to the ease of observing dislocation glide in the fewest 
timesteps as per the timescale of the simulation. However, we noted that 
only the dislocations in the DFT-PFDD simulations remained unpinned and 
mobile after the initial dissociation step shown in 
Fig.~\ref{fig:pfdd-partial-separation}. This is a significant observation 
that underlines the importance of getting the complete shape of the $\gamma$-
surface correctly, especially the barrier heights and shapes. 

The widths of the equilibrium stacking fault were obtained as the separation 
between the peaks on the $\frac{\partial \Delta}{\partial x}$ profiles in 
Fig.~\ref{fig:pfdd-partial-separation}. We only show the SFW obtained from 
the DFT-based PFDD and the MD (GAP) based PFDD because, in the MD (ZW) 
based PFDD results, severe pinning 
did not allow dissociated dislocations to be completely separated, thus 
giving an incorrect ``decrease" in the stacking fault width for a higher Al 
concentration. Another reason for this anomalous decrease in the SFW is 
the decreasing trend in the $C_{44}$ computed using MD (ZW), resulting in  
a decreasing trend in the shear modulus $\mu$ (calculated by Voigt averaging), 
and again for the equilibrium SFW because it is directly proportional to $\mu$ (see Eq.~(\ref{sfw_eq})). 

\begin{figure}[h]
    \centering
    \includegraphics[width=0.95\linewidth]{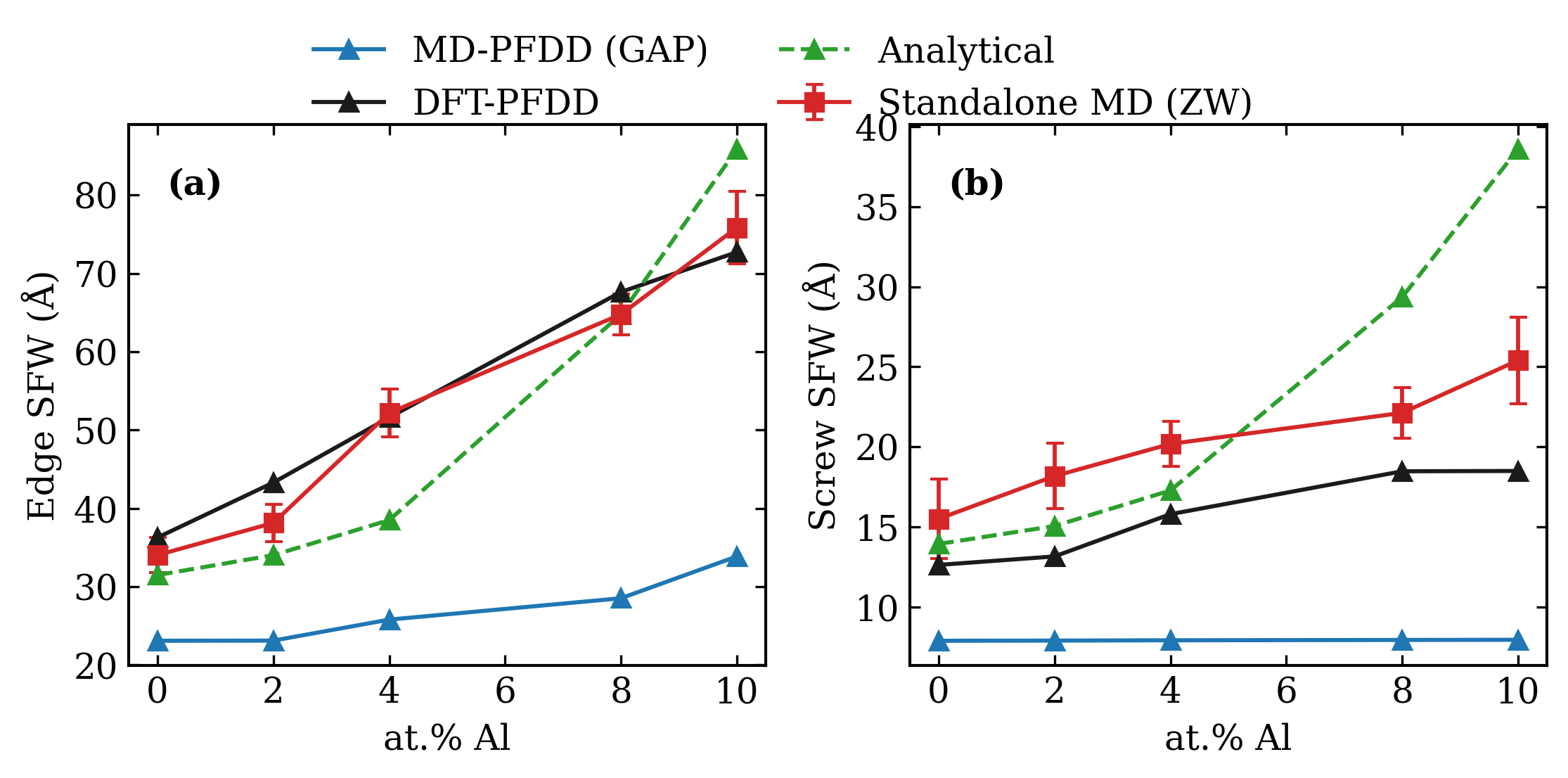}
    \caption{Equilibrium SFW for all alloy cases computed by the multiscale PFDD simulations (DFT-based and MD (GAP)-based) compared with standalone MD (ZW) and analytical SFWs. The lines are meant as a guide to the eye.}
    \label{fig:sfw-full-comp}
\end{figure}

Fig.~\ref{fig:sfw-full-comp} shows the stacking fault widths obtained from
DFT- and MD (GAP)-based PFDD simulations, compared with the standalone MD
simulations and analytically calculated values (Eq.~(\ref{sfw_eq})). The corresponding stacking fault width values for pure-Cu and Cu-Al alloy are summarized in Table~S4 of 
Supplementary Information.
The equilibrium SFWs for edge type dislocations obtained by DFT-PFDD simulations 
show a close agreement with the standalone MD (ZW) results and analytical values. 
The agreement between DFT-PFDD and MD (ZW) is pretty good for the screw case 
as well, even though, the values deviate quite a bit from the analytical 
predictions for larger alloy compositions. This is expected and is attributed to 
the cubic anisotropy of elasticity in Cu-Al system that increases with Al addition, which does 
not get captured in Eq.~(\ref{sfw_eq}). The stacking fault widths obtained from the MD (GAP)-PFDD 
simulations correctly capture the increasing trend in the stacking fault width. But, due 
to the deviation in the shape of the $\gamma$-surface resulting from the higher estimation of the stacking fault energy at the characteristic points, specifically, $T$ and $T_1$.

The agreement between standalone MD (ZW) and DFT-PFDD results is surprising, as the PFDD 
simulations that used MD (ZW)-constructed GSFE surfaces underwent significant dislocation
pinning due to the large deviations in the SFE obtained at the characteristic points, 
especially, $T_1$ and $T$. This strengthens the argument that the design of the interatomic
potentials is an important deciding factor in the success of multiscale PFDD models that 
incorporate GSFE surfaces constructed based on MD simulations. It also shows that while the 
interatomic potential may correctly predict most of the material properties correctly in 
standalone simulations, less-used properties such as the stacking fault energy at the 
characteristic points $T$ and $T_1$ can affect the performance of the potential in a multiscale 
model.

\subsection{MD-PFDD: Scope for improvement}

In PFDD, the evolution of dislocation configurations and the 
dissociation of perfect dislocations into partials, leading 
to the creation of stacking faults, is dictated by the 
balance between elastic energy and the generalised stacking 
fault energy. For this reason, errors in the stacking fault energy 
calculation at the seven significant 
points which capture the shape of the $\gamma$-surface including the points $T$ and $T_1$ can lead to incorrect core structures and improper 
dissociation of dislocations. The MD simulations for $\gamma$-
surface construction relies on interatomic potentials, such 
as ZW potential and, more recently, machine learning potentials 
that learn based on previous DFT simulations. 

Empirical potentials such as ZW potential for Cu-Al are obtained by 
fitting to certain material parameters, such as elastic constants, 
lattice parameters, cohesive energy, and the key stacking fault energy values, 
namely intrinsic SFE (at $G$) and unstable SFE (at $G_2$). From our 
simulations, we see a discrepancy even in the trend of moduli component 
$C_{44}$ calculated by ZW potential even though the lattice parameter 
is predicted accurately by this potential. In addition, due to the
overestimation of the SFE at characteristic points $T$ and $T_1$, 
the curvature of the $\gamma$-surface that has been constructed is also
affected, leading to pinning of the dislocations in the multiscale PFDD model.
Even though the machine learning based GAP potential gets the moduli trend 
correctly, it overestimates the lattice parameter quite similar
to the DFT simulations. This is expected as these ML-based potentials are 
trained based on DFT calculation based datasets. But, contrary to our 
expectation, there is a higher deviation in the GSFE calculated at the seven 
characteristic points as compared to MD (ZW). The deviation in the SFE is most 
significant at $T$ and $T_1$, causing the shape of the GSFE landscape to 
change significantly in this case as well. The separation of the partial 
dislocations after dissociation is incomplete because they are pinned at their 
positions due to the characteristic energy landscape. Through our study, we 
have been able to identify some of the important parameters which the 
interatomic potentials need to correctly capture in order to 
obtain the right equilibrium stacking fault width from a multiscale PFDD 
model. 

\section{Conclusion}{\label{conclusion}}

We have carried out a multiscale investigation of the equilibrium
stacking fault width (SFW) in Cu-Al alloys as a function of Al
content, combining density functional theory (DFT) and molecular
dynamics (MD) using a classical embedded-atom potential (Zhou-Ward-Foster (ZW))
and a machine-learned Gaussian Approximation Potential (GAP), with
phase field dislocation dynamics (PFDD). The atomistic methods
reproduce the lattice parameter and elastic moduli of Cu and Cu-Al
alloys in reasonable agreement with each other and with available
experimental data, establishing that each is individually trustworthy
at the level of standard bulk-property validation. However, when the
generalized stacking fault energy ($\gamma$-surface) from each method
is compared at the seven characteristic points, the MD-derived
surfaces -- for both potentials -- deviate substantially from DFT,
with the largest discrepancies at $T$ and $T_1$. This is striking for
GAP in particular, since it is trained directly on DFT data and might
have been expected to reproduce the DFT $\gamma$-surface more closely
than a classical potential. When incorporated into PFDD, the
DFT-derived $\gamma$-surfaces successfully predict an equilibrium SFW
that increases with Al content for both edge and screw dislocations,
agreeing well with standalone MD measurements and previously reported
experimental values for pure Cu. In contrast, the $\gamma$-surfaces
derived from both MD potentials could not yield a converged
equilibrium SFW within PFDD: the altered landscape causes the
dissociating partials to become pinned soon after separation, rather
than reaching a stable spacing. Surprisingly, despite this failure
within PFDD, the SFW measured directly from standalone MD using
ZW remains in good agreement with the DFT-PFDD predictions,
indicating that the reliability of a $\gamma$-surface for PFDD use is
governed by more than its point-wise accuracy relative to DFT. Taken
together, these results show that neither potential examined here is
currently suitable for driving MD-informed PFDD simulations of
stacking fault behaviour in Cu-Al alloys, and that DFT-derived
$\gamma$-surfaces remain the more reliable, if more expensive, route
to this end. The consistent under-performance of both potentials at
$T$ and $T_1$ suggests that future potential development should
explicitly target the accuracy of the full $\gamma$-surface --
including the shear path sampled during GSFE calculation, not solely
bulk properties -- to enable robust and efficient MD-PFDD modelling of
stacking faults in alloys.

\section*{Credit authorship contribution statement}
\noindent
{\bf Arjun Varma R.:}
PFDD model implementation and simulations; SFW investigation and validation 
using multiscale PFDD model; Visualization; 
Formal analysis; Writing - original 
draft.

\noindent
{\bf Ankosh D. Deshmukh:}
DFT simulations; Investigation of $\gamma$-surface, lattice parameter, and moduli; Visualization; Formal Analysis; Writing - original draft.

\noindent
{\bf Sushil Kumar:}
MD simulations; Investigation of $\gamma$-surface, lattice parameter 
and moduli; Investigation of SFW using standalone MD; Visualization;  Formal Analysis; Writing - original draft.

\noindent
{\bf Vaishali Shah:}
Conceptualization, Investigation of $\gamma$-surface using MACE potential, Formal Analysis, Project administration, Supervision, Writing - review \& editing, Resources \& funding acquisition.

\noindent
{\bf M. P. Gururajan:}
Conceptualization, Formal Analysis, Project administration, Supervision, Writing - review \& editing, Resources \& funding 
acquisition.

\section*{Acknowledgements}
The authors would like to thank the Department of Science and Technology (DST) through research grant CRG/2019/005060 
and National Supercomputing Mission (NSM), India (DST/NSM/R\&D\_HPC\_Applications/2021/03) for the financial support.
We acknowledge use of HPC facilities PARAM Pravega at IISc Bangalore, PARAM
Porul at NIT Trichy, PARAM Siddhi-AI at C-DAC Pune, PARAM Shakti at IIT
Kharagpur under NSM, and Sahasrar at the Department of Scientific Computing
Modeling and Simulation (SCMS), SPPU Pune.
AVR thanks Rahul Chigurupati for the discussions on the implementation of 
the PFDD model. ADD acknowledges the Government of Maharashtra for MAHAJYOTI Fellowship (MAHAJYOTI/Ph.D. Fellowship/2821(28)/2024-25). ADD also thanks Raviraj Mandalia for discussions on slab structures.
The authors acknowledge the use of XcrySDen software~\cite{kokal99} for structure visualization.

\section*{Conflict of interest statement}
The authors have no conflicts to disclose.

\section*{Data availability statement}
The data that support the findings of this study are available from the corresponding author upon reasonable request.

\bibliography{mybibfile}

@article{Thornton1958,
  title={{The effect of stacking fault energy on low temperature creep in pure metals}},
  author={{Thornton, PR and Hirsch, PB}},
  journal={Philosophical Magazine},
  volume={3},
  number={31},
  pages={738--761},
  year={1958},
  publisher={Taylor \& Francis}
}

@article{Whelan1957,
  title={Dislocations and stacking faults in stainless steel},
  author={Whelan, Michael John and Hirsch, Peter Bernhard and Horne, RW and Bollmann, W},
  journal={Proceedings of the Royal Society of London. Series A, Mathematical and Physical Sciences},
  pages={524--538},
  year={1957},
  publisher={JSTOR}
}

@article{Cockayne1971,
  title={The measurement of stacking-fault energies of pure face-centred cubic metals},
  author={Cockayne, D J H and Jenkins, M L and Ray, I L F},
  journal={Philosophical Magazine},
  volume={24},
  number={192},
  pages={1383--1392},
  year={1971},
  publisher={Taylor \& Francis}
}

@article{Shi2024,
  title={In-situ investigation of the interaction between hydrogen and stacking faults in a bulk austenitic steel},
  author={Shi, Hao and Nandy, Supriya and Cheng, Huijie and Sun, Binhan and Ponge, Dirk},
  journal={Acta Materialia},
  volume={262},
  pages={119441},
  year={2024},
  publisher={Elsevier}
}

@article{Yu2025,
  title={{The effects of alloying elements on hydrogen-affected generalized stacking fault energy and dislocation behavior in Nickel-based alloys}},
  author={Yu, Chenjun and Okita, Taira and Kawabata, Tomoya and Uranaka, Shohei and Li, Xixian},
  journal={Materials Chemistry and Physics},
  volume={341},
  pages={130907},
  year={2025},
  publisher={Elsevier}
}

@article{Xie2021,
  title={Sluggish hydrogen diffusion and hydrogen decreasing stacking fault energy in a high-entropy alloy},
  author={Xie, Zhoucan and Wang, Yunjiang and Lu, Chunsheng and Dai, Lanhong},
  journal={Materials Today Communications},
  volume={26},
  pages={101902},
  year={2021},
  publisher={Elsevier}
}

@article{Hakkinen1990,
  title={{Edge dislocations in fcc metals: Microscopic calculations of core structure and positron states in Al and Cu}},
  author={H{\"a}kkinen, H and M{\"a}kinen, S and Manninen, M},
  journal={Physical Review B},
  volume={41},
  number={18},
  pages={12441},
  year={1990},
  publisher={APS}
}

@article{Srinivasan2005,
  title={Compact and dissociated dislocations in aluminum: implications for deformation},
  author={Srinivasan, SG and Liao, XZ and Baskes, MI and McCabe, RJ and Zhao, YH and Zhu, YT},
  journal={Physical review letters},
  volume={94},
  number={12},
  pages={125502},
  year={2005},
  publisher={APS}
}

@article{Tsuzuki2009,
  title={Molecular dynamics simulation of fast dislocations in copper},
  author={Tsuzuki, H{\'e}lio and Branicio, Paulo Sergio and Rino, Jos{\'e} Pedro},
  journal={Acta Materialia},
  volume={57},
  number={6},
  pages={1843--1855},
  year={2009},
  publisher={Elsevier}
}

@article{Aslanides1998,
  title={Atomistic study of dislocation cores in aluminium and copper},
  author={Aslanides, A and Pontikis, V},
  journal={Computational materials science},
  volume={10},
  number={1-4},
  pages={401--405},
  year={1998},
  publisher={Elsevier}
}

@article{Henager2005,
  title={Dislocation and stacking fault core fields in fcc metals},
  author={Henager Jr, CH and Hoagland, Richard G},
  journal={Philosophical Magazine},
  volume={85},
  number={36},
  pages={4477--4508},
  year={2005},
  publisher={Taylor \& Francis}
}

@article{Yi2017,
  title={Solute softening and defect generation during prismatic slip in magnesium alloys},
  author={Yi, Peng and Cammarata, Robert C and Falk, Michael L},
  journal={Modelling and Simulation in Materials Science and Engineering},
  volume={25},
  number={8},
  pages={085001},
  year={2017},
  publisher={IOP Publishing}
}

@article{Rao2017,
  title={Atomistic simulations of dislocation behavior in a model FCC multicomponent concentrated solid solution alloy},
  author={Rao, Satish I and Woodward, Christopher and Parthasarathy, Triplicane A and Senkov, Oleg},
  journal={Acta Materialia},
  volume={134},
  pages={188--194},
  year={2017},
  publisher={Elsevier}
}

@article{Shih2021,
  title={Stacking fault energy in concentrated alloys},
  author={Shih, Mulaine and Miao, Jiashi and Mills, Michael and Ghazisaeidi, Maryam},
  journal={Nature communications},
  volume={12},
  number={1},
  pages={3590},
  year={2021},
  publisher={Nature Publishing Group UK London}
}

@article{Wise2025,
  title={Cross slip of extended dislocations in face-centered cubic metals through phase-field modeling},
  author={Wise, Ian and Hunter, Abigail and Beyerlein, Irene J},
  journal={Modelling and Simulation in Materials Science and Engineering},
  volume={33},
  number={5},
  pages={055011},
  year={2025},
  publisher={IOP Publishing}
}

@article{Xu2022,
  title={Dislocation dynamics in heterogeneous nanostructured materials},
  author={Xu, Shuozhi and Cheng, Justin Y and Mara, Nathan A and Beyerlein, Irene J},
  journal={Journal of the Mechanics and Physics of Solids},
  volume={168},
  pages={105031},
  year={2022},
  publisher={Elsevier}
}

@article{Zheng2023,
  title={{Multi-scale investigation of short-range order and dislocation glide in MoNbTi and TaNbTi multi-principal element alloys}},
  author={Zheng, Hui and Fey, Lauren TW and Li, Xiang-Guo and Hu, Yong-Jie and Qi, Liang and Chen, Chi and Xu, Shuozhi and Beyerlein, Irene J and Ong, Shyue Ping},
  journal={npj Computational Materials},
  volume={9},
  number={1},
  pages={89},
  year={2023},
  publisher={Nature Publishing Group UK London}
}

@article{Peng2023,
  title={{A combined kinetic Monte Carlo and phase field approach to model thermally activated dislocation motion}},
  author={Peng, Xiaoyao and Mathew, Nithin and Beyerlein, Irene J and Martinez, Enrique and Hunter, Abigail},
  journal={Computational Materials Science},
  volume={230},
  pages={112490},
  year={2023},
  publisher={Elsevier}
}

@article{Kim1985,
  title={{Mechanical and kinetic effects of cathodic hydrogen charging of copper and of Cu-Al alloys}},
  author={Kim, Jong Jip and Byrne, JG},
  journal={Materials Science and Engineering},
  volume={74},
  number={2},
  pages={201--214},
  year={1985},
  publisher={Elsevier}
}

@article{Pan1986,
  title={{Low temperature behavior of hydrogen in copper and Cu-Al alloys}},
  author={Pan, Yi and Byrne, JG},
  journal={Materials Science and Engineering},
  volume={84},
  pages={195--204},
  year={1986},
  publisher={Elsevier}
}

@article{Rohatgi2001,
  title={The influence of stacking fault energy on the mechanical behavior of {Cu} and {Cu-Al} alloys: deformation twinning, work hardening, and dynamic recovery},
  author={Rohatgi, Aashish and Vecchio, Kenneth S and Gray III, George T},
  journal={Metallurgical and Materials Transactions A},
  volume={32},
  number={1},
  pages={135--145},
  year={2001},
  publisher={Springer}
}

@article{Sandlobes2012,
  title={{The relation between ductility and stacking fault energies in Mg and Mg--Y alloys}},
  author={Sandl{\"o}bes, S and Fri{\'a}k, M and Zaefferer, S and Dick, A and Yi, Sang-bong and Letzig, D and Pei, Zong-rui and Zhu, L-F and Neugebauer, J and Raabe, D},
  journal={Acta Materialia},
  volume={60},
  number={6-7},
  pages={3011--3021},
  year={2012},
  publisher={Elsevier}
}

@article{Youssef2011,
  title={{Effect of stacking fault energy on mechanical behavior of bulk nanocrystalline Cu and Cu alloys}},
  author={Youssef, Khaled and Sakaliyska, Miroslava and Bahmanpour, Hamed and Scattergood, Ronald and Koch, Carl},
  journal={Acta Materialia},
  volume={59},
  number={14},
  pages={5758--5764},
  year={2011},
  publisher={Elsevier}
}

@article{Pierce2015,
  title={{The influence of stacking fault energy on the microstructural and strain-hardening evolution of Fe--Mn--Al--Si steels during tensile deformation}},
  author={Pierce, Donald T and Jim{\'e}nez, Jose Antonio and Bentley, James and Raabe, Dierk and Wittig, James E},
  journal={Acta Materialia},
  volume={100},
  pages={178--190},
  year={2015},
  publisher={Elsevier}
}

@article{Yin2016,
  title={{Mechanical behavior and deformation kinetics of gradient structured Cu-Al alloys with varying stacking fault energy}},
  author={Yin, Zhe and Sun, Lele and Yang, Jian and Gong, Yulan and Zhu, Xinkun},
  journal={Journal of Alloys and Compounds},
  volume={687},
  pages={152--160},
  year={2016},
  publisher={Elsevier}
}

@article{Mahato2015,
  title={{An effective stacking fault energy viewpoint on the formation of extended defects and their contribution to strain hardening in a Fe--Mn--Si--Al twinning-induced plasticity steel}},
  author={Mahato, B and Shee, SK and Sahu, T and Chowdhury, S Ghosh and Sahu, P and Porter, DA and Karjalainen, LP},
  journal={Acta Materialia},
  volume={86},
  pages={69--79},
  year={2015},
  publisher={Elsevier}
}

@article{perdew96,
  title = {Generalized {G}radient {A}pproximation {M}ade {S}imple},
  author = {Perdew, John P. and Burke, Kieron and Ernzerhof, Matthias},
  journal = {Phys. Rev. Lett.},
  volume = {77},
  issue = {18},
  pages = {3865--3868},
  numpages = {0},
  year = {1996},
  month = {Oct},
  publisher = {American Physical Society},
}

@article{ravindran98,
author = {Ravindran,P.  and Fast,Lars  and Korzhavyi,P. A.  and Johansson,B.  and Wills,J.  and Eriksson,O. },
title = {Density functional theory for calculation of elastic properties of orthorhombic crystals: {A}pplication to {TiSi$_2$}},
journal = {Journal of Applied Physics},
volume = {84},
number = {9},
pages = {4891-4904},
year = {1998},
}

@article{plimpton1995,
title = {Fast Parallel Algorithms for Short-Range Molecular Dynamics},
author = {Steve Plimpton},
journal = {Journal of Computational Physics},
volume = {117},
number = {1},
pages = {1-19},
year = {1995},
issn = {0021-9991}
}

@article{kresse96a,
  title = {Efficient iterative schemes for ab initio total{-}energy calculations using a plane{-}wave basis set},
  author = {Kresse, G. and Furthm{\"u}ller, J.},
  journal = {Phys. Rev. B},
  volume = {54},
  issue = {16},
  pages = {11169--11186},
  numpages = {0},
  year = {1996},
  month = {Oct},
  publisher = {American Physical Society},

}

@article{kresse96b,
title = {Efficiency of ab{-}initio total energy calculations for metals and semiconductors using a plane{-}wave basis set},
journal = {Computational Materials Science},
volume = {6},
number = {1},
pages = {15-50},
year = {1996},
issn = {0927-0256},
author = {G. Kresse and J. Furthm{\"u}ller}
}

@article{blochl94,
  title = {Projector augmented{-}wave method},
  author = {Bl{\"o}chl, P. E.},
  journal = {Phys. Rev. B},
  volume = {50},
  issue = {24},
  pages = {17953--17979},
  numpages = {0},
  year = {1994},
  month = {Dec},
  publisher = {American Physical Society},
}

@article{kresse99,
  title = {From ultrasoft pseudopotentials to the projector augmented{-}wave method},
  author = {Kresse, G. and Joubert, D.},
  journal = {Phys. Rev. B},
  volume = {59},
  issue = {3},
  pages = {1758--1775},
  numpages = {0},
  year = {1999},
  month = {Jan},
  publisher = {American Physical Society},
}

@article{Hunter2011,
  title = {Influence of the stacking fault energy surface on partial dislocations in fcc metals with a three{-}dimensional phase field dislocations dynamics model},
  author = {Hunter, A. and Beyerlein, I. J. and Germann, T. C. and Koslowski, M.},
  journal = {Phys. Rev. B},
  volume = {84},
  issue = {14},
  pages = {144108},
  numpages = {10},
  year = {2011},
  month = {Oct},
  publisher = {American Physical Society},
}

@article{monkhorst76,
  title = {Special points for {B}rillouin-zone integrations},
  author = {Monkhorst, Hendrik J. and Pack, James D.},
  journal = {Phys. Rev. B},
  volume = {13},
  issue = {12},
  pages = {5188-5192},
  numpages = {0},
  year = {1976},
  month = {Jun},
  publisher = {American Physical Society},
}

@article{methfessel89,
  title = {High{-}precision sampling for {B}rillouin{-}zone integration in metals},
  author = {Methfessel, M. and Paxton, A. T.},
  journal = {Phys. Rev. B},
  volume = {40},
  issue = {6},
  pages = {3616--3621},
  numpages = {0},
  year = {1989},
  month = {Aug},
  publisher = {American Physical Society},
}

@article{vitek68,
author = { V. Vítek },
title = {Intrinsic stacking faults in body{-}centred cubic crystals},
journal = {The Philosophical Magazine: A Journal of Theoretical Experimental and Applied Physics},
volume = {18},
number = {154},
pages = {773-786},
year  = {1968},
publisher = {Taylor & Francis},
}

@article{kokal99,
title = {{XCrySDen}{—}a new program for displaying crystalline structures and electron densities},
journal = {Journal of Molecular Graphics and Modelling},
volume = {17},
number = {3},
pages = {176-179},
year = {1999},
issn = {1093-3263},
author = {Anton Kokalj}
}

@article{Wang2010,
	author = {Wang, Yunzhi and Li, Ju},
	journal = {Acta Materialia},
	number = {4},
	pages = {1212--1235},
	publisher = {Acta Materialia Inc.},
	title = {{Phase field modeling of defects and 
	deformation}},
	volume = {58},
	year = {2010}
}

@article{Beyerlein2016,
	author = {Beyerlein, I J and Hunter, A},
	journal = {Philosophical Transactions of 
	Royal Society A},
	number = {2066},
	pages = {{20150166}},
	title = {{Understanding nanoscale dislocation 
	mechanics using phase field dislocation 
	dynamics}},
	volume = {374},
	year = {2016}
}

@article{Hunter2011a,
	journal = {International Journal of High 
	Performance Computing Applications},
	number = {2},
	pages = {223--235},
	title = {{Large-scale 3D phase field 
	dislocation dynamics simulations on 
	high-performance architectures}},
	author = {Hunter, A and Beyerlein, I J},
	volume = {25},
	year = {2011}
}

@article{Hunter2014,
	author = {Hunter, A and Beyerlein, I J},
	journal = {Materials Science {\&} Engineering 
	A},
	pages = {200--210},
	title = {{Stacking fault emission from grain 
	boundaries : Material dependencies and grain 
	size effects}},
	volume = {600},
	year = {2014}
}

@article{Shen2004,
	author = {Shen, C. and Wang, Y.},
	journal = {Acta Materialia},
	number = {3},
	pages = {683--691},
	title = {{Incorporation of $\gamma$-surface 
	to phase field model of dislocations: 
	Simulating dislocation dissociation in fcc 
	crystals}},
	volume = {52},
	year = {2004}
}

@article{Shen2014,
	author = {Shen, Chen and Li, Ju and Wang, 
	Yunzhi},
	journal = {Acta Materialia},
	pages = {125--131},
	title = {{Predicting structure and energy of 
	dislocations and grain boundaries}},
	volume = {74},
	year = {2014}
}

@article{Hunter2015,
	author = {Hunter, A. and Beyerlein, I. J.},
	journal = {Acta Materialia},
	pages = {207--217},
	title = {{Relationship between monolayer 
	stacking faults and twins in nanocrystals}},
	volume = {88},
	year = {2015}
}

@book{Mura1987,
  title={Micromechanics of defects in solids},
  author={Mura, T},
  year={1987},
  publisher={Martinus Nijhoff Publishers}
}

@article{Schoeck2001,
  title={{The core structure, recombination 
  energy and Peierls energy for dislocations in 
  Al}},
  author={Schoeck, Gunther},
  journal={Philosophical Magazine A},
  volume={81},
  number={5},
  pages={1161--1176},
  year={2001},
  publisher={Taylor \& Francis}
}

@article{Kamalakshi2022,
  title={{Deformation behaviour of Cu and Cu--Al 
  in the dislocation starved regime: A molecular 
  dynamics study}},
  author={Kamalakshi, G and Pant, Prita and 
  Gururajan, M P},
  journal={Computational Materials Science},
  volume={203},
  pages={111087},
  year={2022},
  publisher={Elsevier}
}

@article{Stobbs1971,
	title={{The weak beam technique as applied to 
	the determination of the stacking-fault 
	energy of Copper}},
	author={Stobbs, W M and Sworn, C H},
	journal={Philosophical Magazine},
	volume={24},
	number={192},
	pages={1365--1381},
	year={1971},
	publisher={Taylor \& Francis}
}

@article{Shen2003,
  title={Phase field model of dislocation networks},
  author={Shen, C and Wang, Y},
  journal={Acta Materialia},
  volume={51},
  number={9},
  pages={2595--2610},
  year={2003},
  publisher={Elsevier}
}

@article{Varma2023,
  title={Dislocation assisted phase separation: A 
  phase field study},
  author={Varma R., Arjun and Pant, Prita and 
  Gururajan, M P},
  journal={Acta Materialia},
  volume={244},
  pages={118529},
  year={2023},
  publisher={Elsevier}
}

@article{Szajewski2018,
  title={The influence of anisotropy on the core structure of Shockley partial dislocations within FCC materials},
  author={Szajewski, BA and Hunter, A and Luscher, DJ and Beyerlein, IJ},
  journal={Modelling and Simulation in Materials Science and Engineering},
  volume={26},
  number={1},
  pages={015010},
  year={2018},
  publisher={IOP Publishing}
}

@article{cain1971,
  title = {Elastic Constants of ${\alpha}-$Phase {Cu-Al} Alloys},
  author = {Cain, L. S. and Thomas, J. F.},
  journal = {Phys. Rev. B},
  volume = {4},
  issue = {12},
  pages = {4245--4255},
  numpages = {0},
  year = {1971},
  month = {Dec},
  publisher = {American Physical Society},
}

@article{Liu1999,
  title={A new investigation of copper's role in enhancing Al--Cu interconnect electromigration resistance from an atomistic view},
  author={Liu, X-Y and Liu, C-L and Borucki, LJ},
  journal={Acta materialia},
  volume={47},
  number={11},
  pages={3227--3231},
  year={1999},
  publisher={Elsevier}
}

@article{Zhou2004,
  title={{Misfit-energy-increasing dislocations in vapor-deposited CoFe/NiFe multilayers}},
  author={Zhou, Xiaowang W and Johnson, RA and Wadley, Haydn NG},
  journal={Physical Review B},
  volume={69},
  number={14},
  pages={144113},
  year={2004},
  publisher={APS}
}

@article{Ward2012,
  title={{Rapid production of accurate embedded-atom method potentials for metal alloys}},
  author={Ward, Logan and Agrawal, Anupriya and Flores, Katharine M and Windl, Wolfgang},
  journal={arXiv preprint arXiv:1209.0619},
  year={2012}
}

@misc{note1,
title = {These potentials gave a better agreement of both lattice parameter and elastic constants, whereas, simulations using {PAW} potentials with valence electronic configuration of {3$p^6$ s$d^{10}$ 4s$^1$} for {Cu} gave poor agreement of elastic constants with experimental data.}
}

@misc{note2,
title = {Simulations with l6 layers slabs with stacking fault between layers 8 and 9, gave qualitatively same results as 10 layers slabs. {T}he results of 10 layer structures show a better agreement with published data, hence, to keep the computational costs optimal, all results were obtained with 10 layer structures. {H}owever, the agreement of {G} point stacking fault energy with 16 layer slabs is better than 10 layer slabs with experimental data.}
}

@article{mace,     
    author = {Batatia, Ilyes and Benner, Philipp and Chiang, Yuan and Elena, Alin M. and Kovács, Dávid P. and Riebesell, Janosh and Advincula, Xavier R. and Asta, Mark and Avaylon, Matthew and Baldwin, William J. and Berger, Fabian and Bernstein, Noam and Bhowmik, Arghya and Bigi, Filippo and Blau, Samuel M. and Cărare, Vlad and Ceriotti, Michele and Chong, Sanggyu and Darby, James P. and De, Sandip and Della Pia, Flaviano and Deringer, Volker L. and Elijošius, Rokas and El-Machachi, Zakariya and Fako, Edvin and Falcioni, Fabio and Ferrari, Andrea C. and Gardner, John L. A. and Gawkowski, Mikołaj J. and Genreith-Schriever, Annalena and George, Janine and Goodall, Rhys E. A. and Grandel, Jonas and Grey, Clare P. and Grigorev, Petr and Han, Shuang and Handley, Will and Heenen, Hendrik H. and Hermansson, Kersti and Ho, Cheuk Hin and Hofmann, Stephan and Holm, Christian and Jaafar, Jad and Jakob, Konstantin S. and Jung, Hyunwook and Kapil, Venkat and Kaplan, Aaron D. and Karimitari, Nima and Kermode, James R. and Kourtis, Panagiotis and Kroupa, Namu and Kullgren, Jolla and Kuner, Matthew C. and Kuryla, Domantas and Liepuoniute, Guoda and Lin, Chen and Margraf, Johannes T. and Magdău, Ioan-Bogdan and Michaelides, Angelos and Moore, J. Harry and Naik, Aakash A. and Niblett, Samuel P. and Norwood, Sam Walton and O’Neill, Niamh and Ortner, Christoph and Persson, Kristin A. and Reuter, Karsten and Rosen, Andrew S. and Rosset, Louise A. M. and Schaaf, Lars L. and Schran, Christoph and Shi, Benjamin X. and Sivonxay, Eric and Stenczel, Tamás K. and Sutton, Christopher and Svahn, Viktor and Swinburne, Thomas D. and Tilly, Jules and van der Oord, Cas and Vargas, Santiago and Varga-Umbrich, Eszter and Vegge, Tejs and Vondrák, Martin and Wang, Yangshuai and Witt, William C. and Wolf, Thomas and Zills, Fabian and Csányi, Gábor},
  title = {A foundation model for atomistic materials chemistry},
    journal = {The Journal of Chemical Physics},
    volume = {163},
    number = {18},
    pages = {184110},
    year = {2025},   
    issn = {0021-9606},
    month = {11},
}

@article{parrinello,
  title = {Generalized Neural-Network Representation of High-Dimensional Potential-Energy Surfaces},
  author = {Behler, J\"org and Parrinello, Michele},
  journal = {Phys. Rev. Lett.},
  volume = {98},
  issue = {14},
  pages = {146401}, 
  numpages = {4},
  year = {2007},
  month = {Apr},
  publisher = {American Physical Society},
}

@article{gap,
  title = {Gaussian Approximation Potentials: The Accuracy of Quantum Mechanics, without the Electrons},
  author = {Bart\'ok, Albert P. and Payne, Mike C. and Kondor, Risi and Cs\'anyi, G\'abor},
  journal = {Phys. Rev. Lett.},
  volume = {104},
  issue = {13},
  pages = {136403},
  numpages = {4},
  year = {2010},
  month = {Apr},
  publisher = {American Physical Society},
}

@article{soap,
  title = {On representing chemical environments},
  author = {Bart\'ok, Albert P. and Kondor, Risi and Cs\'anyi, G\'abor},
  journal = {Phys. Rev. B},
  volume = {87},
  issue = {18},
  pages = {184115},
  numpages = {16},
  year = {2013},
  month = {May},
  publisher = {American Physical Society},

}

@article{mishin,
title = {Machine-learning interatomic potentials for materials science},
journal = {Acta Materialia},
volume = {214},
pages = {116980},
year = {2021},
issn = {1359-6454},
author = {Y. Mishin},
}

@article{Antillon2025,
author = {Antillon, Edwin A},
journal = {Physical Review Materials},
number = {8},
pages = {83801},
publisher = {American Physical Society},
title = {{Thermodynamical phase-stability of a Cu-Al binary system using machine-learning interatomic potentials}},
volume = {9},
year = {2025}
}

@article{Moment1972,
  title={Elastic stiffnesses of copper-tin and copper-aluminum alloy single crystals},
  author={Moment, Roger L},
  journal={Journal of Applied Physics},
  volume={43},
  number={11},
  pages={4419--4424},
  year={1972},
  publisher={American Institute of Physics}
}

\pagebreak
%
%
%
%
%

\renewcommand{\thefootnote}{\fnsymbol{footnote}}
\setcounter{secnumdepth}{5}

\begin{center}{\LARGE \bf \underline{Supplementary Information}}\end{center}
\vspace{10pt}

\begin{center}
{\LARGE \bf DFT- versus MD-derived Generalized Stacking Fault Energy Surfaces for Phase-Field Dislocation Dynamics: Multiscale Prediction of Stacking Fault Widths in Cu–Al Alloys }
\end{center}

\begin{figure}[h]
\begin{center}
\includegraphics[scale=0.2]{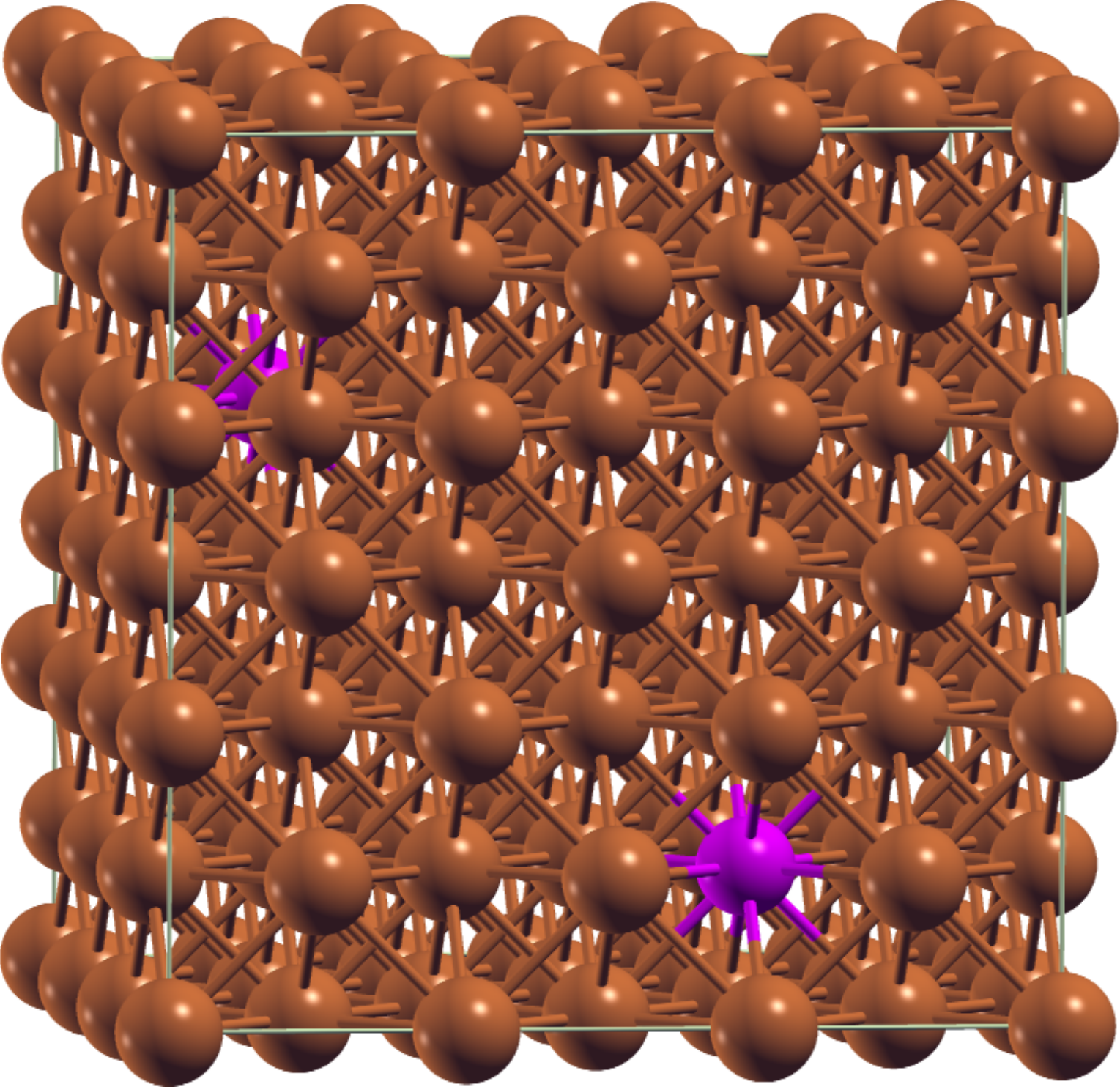} \\ \vskip 0.1 in
Figure S1: Optimized $3\times3\times3$ unit cell crystal structure of Cu2Al alloy. Atoms in blue and pink color are Cu and Al, respectively.
\label{bulk}
\end{center}
\end{figure}

\begin{figure}[h]
\begin{center}
\includegraphics[scale=0.5]{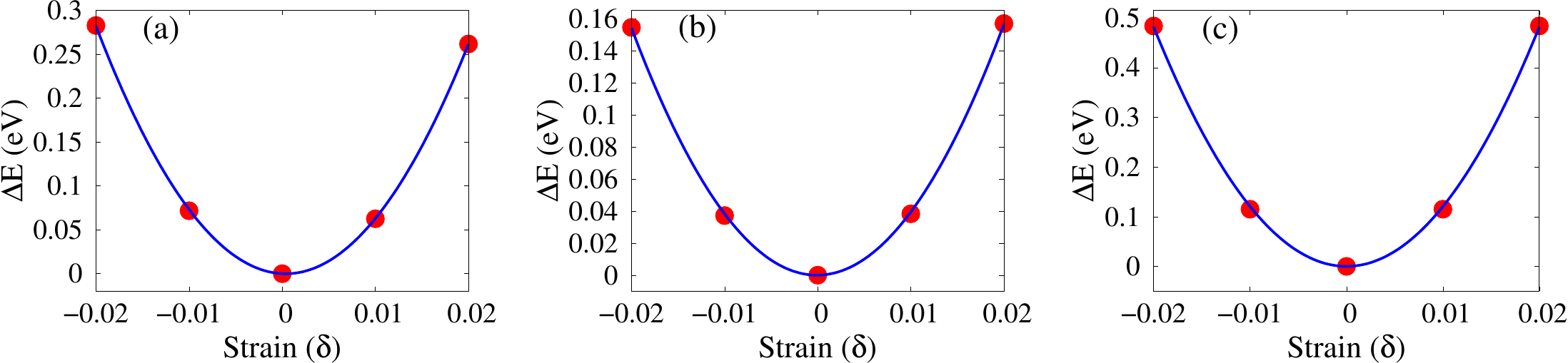} \\
Figure S2: Energy change, $\Delta E=E(V,\delta)-E(V_0,0)$, as a function of strain for determining (a) $C_{11}$, (b) $C_{12}$, and (c) $C_{44}$ of Cu2Al. Solid curve represents polynomial fit using Eq.~(1) of the main text for the corresponding elastic constant.
\label{elast}
\end{center}
\end{figure}


\begin{figure}[h]
\begin{center}
\includegraphics[scale=0.5]{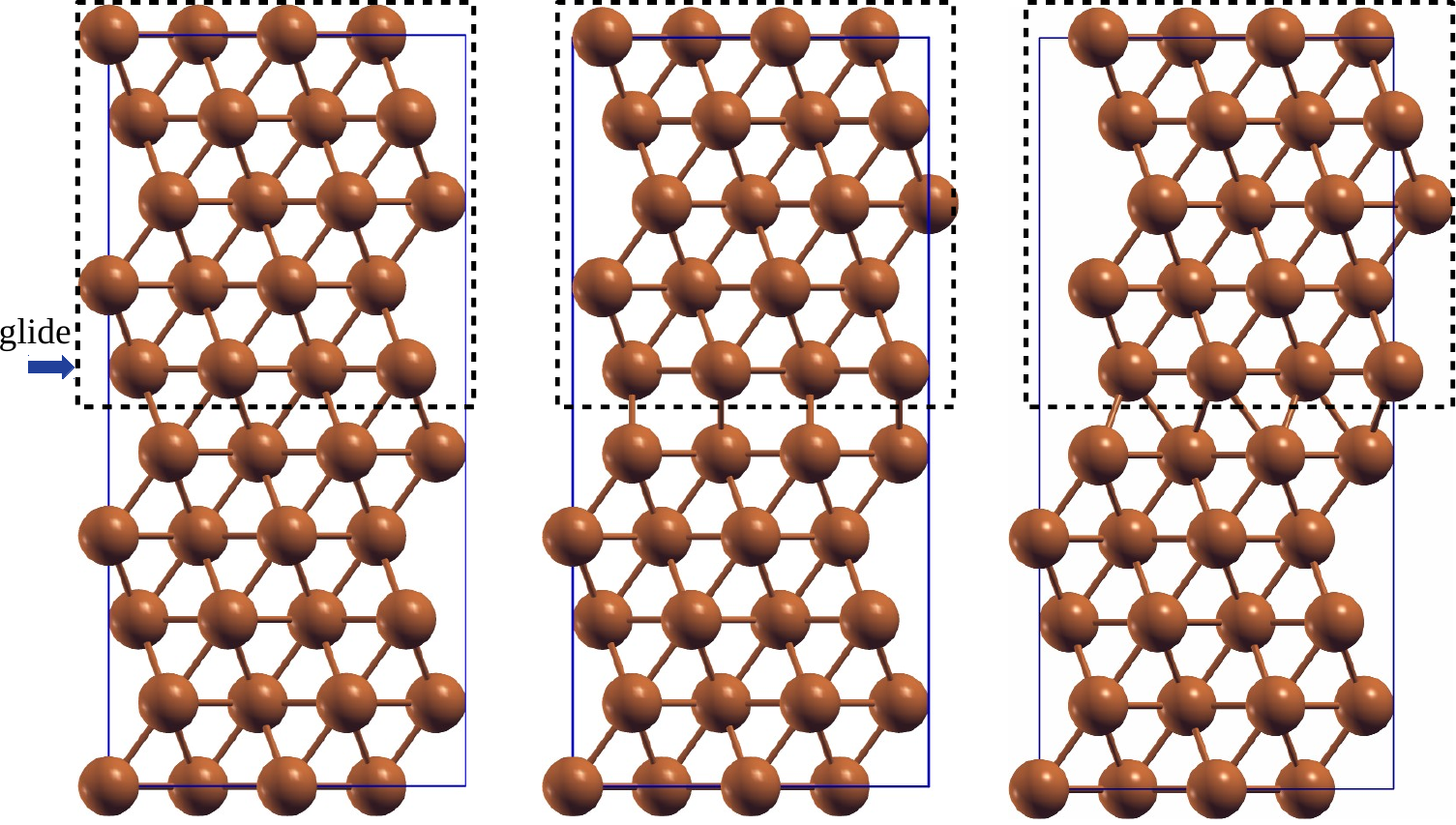} \\
\hskip 0.15 in  (a) \hskip 1.4 in (b) \hskip 1.4 in (c) \\
Figure S3: Front view of the Cu crystal slab (a) without and with rigid displacements for special points (b) G$_2$ and (c) G.
For clariry, the structures are shown without vacuum; however, a vacuum spacing of 15 \AA\ was added in the actual simulation cell. Atoms in blue and pink color are Cu and Al, respectively.
\label{cu-un}
\end{center}
\end{figure}

\begin{figure}[h]
\begin{center}
\includegraphics[scale=1.0]{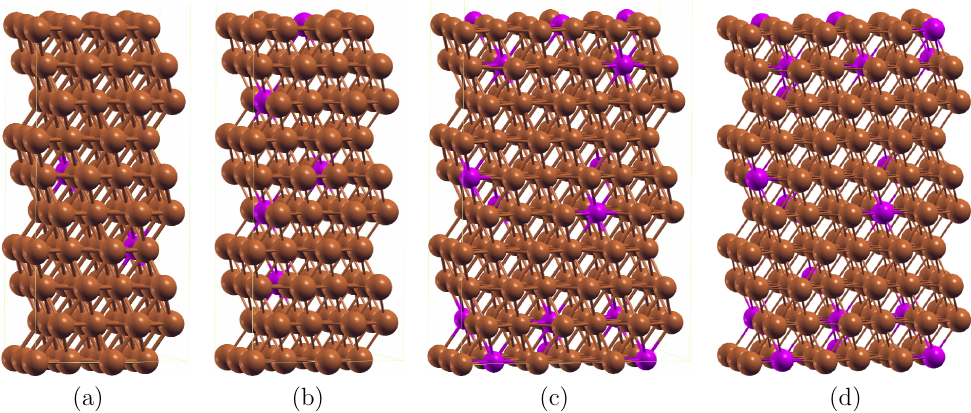} \\
Figure S4: Crystal structure of Cu$x$Al alloy for (a) $x=$2, (b) $x=$4, (c) $x=$8, and (d) $x=$10. Atoms in blue and pink color are Cu and Al, respectively.
\label{cu-un-alloy}
\end{center}
\end{figure}

\begin{table}[h]
\centering
Table S1: {Number of Al atoms in the Cu$x$Al ($x$=2, 4, 8, 10) alloy with the layerwise distribution in the stacking fault energy calculations. All Al atoms are randomly substituted at the Cu sites within and across the respective Cu layers.} \\ \vskip 0.1 in
    \begin{tabular}{cccc}
    \hline
   Material & $\#$ Al atoms & layer $\#$ & $\#$ Al atom(s) in layer \\
    \hline
    Cu2Al &   2 &    4,6 &  1,1  \\
    Cu4Al &   5 &    3,5,6,8,10  & 1,1,1,1,1\\
    Cu8Al &   15 &   1,2,5,6,9,10 & 3,3,2,2,2,3 \\
    Cu10Al &  18 &  1,2,3,5,6,8,9,10 & 3,3,1,2,2,1,3,3\\
    \hline
    \end{tabular}
    \label{Alatoms}
\end{table}

\begin{table}[h]
    \centering
    Table S2: {The elastic moduli (in GPa) and lattice parameter $a$ (\AA) values calculated for pure Cu and Cu-Al alloys using both DFT and MD methods.} \\ \vskip 0.1 in
    \scalebox{0.83}{
    \begin{tabular}{lccccccccccccccccccc}
        \hline
       \multirow{2}{*}{Material} & \multicolumn{4}{c}{DFT} & &\multicolumn{4}{c}{MD (ZW)} & &\multicolumn{4}{c}{MD (GAP)} && \multicolumn{4}{c}{MD (MACE)}  \\
        \cline{2-5} \cline{7-10} \cline{12-15}  \cline{17-20}
                      & $a$  & $C_{11}$ & $C_{12}$ & $C_{44}$ & & $a$  & $C_{11}$ & $C_{12}$ & $C_{44}$ &
                      & $a$  & $C_{11}$ & $C_{12}$ & $C_{44}$ &
                      & $a$  & $C_{11}$ & $C_{12}$ & $C_{44}$\\
        \hline
        Cu    & 3.635 & 169.908 & 121.030 & 75.059 &&
                3.614 & 170.77  & 121.35  & 75.98  &&
                3.632 & 168.1   & 119.5   & 74.7 &&
                3.635 & 168.7   & 126.6   & 75.5 \\
        Cu2Al & 3.639 & 167.485 & 119.677 & 74.234  &&
                3.619 & 167.98  & 121.56  & 73.36 &&
                3.636 & 166.3   & 118.0   & 75.4 &&
                3.639 & 166.8   & 125.4   & 74.3 \\
        Cu4Al & 3.643 & 169.449 & 119.659 & 76.437 &&
                3.626 & 163.73  & 121.37  & 69.88 && 
                3.643 & 164.8   & 117.5   & 76.2 &&
                3.643 & 164.3   & 123.4   & 72.6 \\
        Cu8Al & 3.654 & 165.389 & 117.196 & 76.702 &&
                3.639 & 156.42  & 119.87  & 64.71 &&
                3.654 & 162     & 116.7   & 76.7  &&
                3.654 & 160.5   & 120.1   & 68.7 \\
       Cu10Al & 3.658 & 165.600 & 119.200 & 76.870 &&
                3.644 & 153.24  & 118.81  & 62.78 &&
                3.658 & 160.7   & 116.4   & 76.6 &&
                3.658 & 157.8   & 118.3   & 67.2 \\
        \hline
    \end{tabular}
    }
    \label{tab:moduli_latparam}
\end{table}

\begin{table}[h]
   \centering
 Table S3: {The stacking fault energy values (in $mJ/m^2$) at the seven  points labelled on the GSFE surface (see section~2.1.2) and calculated using DFT and MD simulations.} \\ \vskip 0.1 in
    \begin{tabular}{lllllllll}
    \hline
    Material & Sim. type & $A$ & $G_1$ & $G_2$ & $G_3$ & $G$ & $T_1$ & $T$ \\
    \hline   
    Cu & DFT & 0 & 82.66 & 156.27 & 108.37 & 38.85 & 288.53 &  513.99 \\ 
     & MD (ZW) & 0 & 83.76 & 178.95 & 112.95 & 40.21 & 344.68  & 710.03 \\
     & MD (GAP) & 0 & 93.33 & 185.76 & 125.53 & 40.26 & 358.88  & 756.00 \\
     & MD (MACE) & 0 & 78.37 & 155.19 & 95.15 & 17.79 & 317.54  & 643.86 \\
     \hline
    Cu2Al & DFT & 0 & 83.79 & 153.63 & 98.37 & 23.43 & 284.76 & 499.33 \\ 
     & MD (ZW) & 0 & 79.65 & 170.85 & 106.25 & 35.90 & 334.41 & 696.11 \\
     & MD (GAP) & 0 & 92.49 & 183.31 & 121.90 & 35.87 & 356.03 & 751.01 \\
     & MD (MACE) & 0 & 76.43 & 151.56 & 93.26 & 17.88 & 310.41 & 631.91 \\ 
      \hline
    Cu4Al & DFT & 0 & 83.75 & 152.05 & 93.38 & 13 & 283.82 & 497.36 \\ 
     & MD (ZW) & 0 & 74.05 & 159.38 & 96.63 & 29.94 & {325.43} &{675.44} \\ 
     & MD (GAP) & 0 & 90.61 & 177.85 & 114.10 & 28.22 & 349.49 & 738.32 \\
     & MD (MACE) & 0 & 76.01 & 150.73 & 92.77 & 17.71 & 308.94 & 630.65 \\
     \hline
    Cu8Al & DFT & 0 & 81.03 & 146.97 & 88.10 & 9.92 & 275.66 & 482.25 \\ 
     & MD (ZW) & 0 & 66.94 & 143.20 & 79.89 & 16.37 & 304.00 & 649.97 \\
     & MD (GAP) & 0 & 88.27 & 172.07 & 107.08 & 19.45 & 341.31 & 725.11 \\
     & MD (MACE) & 0 & 68.67 & 130.73 & 68.00 & -     & 289.28 & 581.96 \\
     \hline
    Cu10Al & DFT & 0 & 79.79 & 144.43 & 86.00 & 8.63 & 273.28 & 478.81 \\
     & MD (ZW) & 0 & 65.59 & 140.12 & 76.23 & 12.03 & 301.04 & 645.69 \\
     & MD (GAP) & 0 & 86.21 & 166.06 & 99.16 & 12.80 & 336.33 & 715.03 \\
     & MD (MACE) & 0 & 71.17 & 142.99 & 94.97 & 42.94 & 286.79  & 602.42 \\
    \hline
    \end{tabular}
    \label{tab:gsfe-points-dft}
\end{table}

\begin{table}[h!]
    \centering 
  Table S4: {Widths of stacking faults formed between edge and screw dislocations from DFT-PFDD simulations and standalone MD simulations.}  \\ \vskip 0.1 in
\begin{tabular}{lcccccccc}
        \hline
      \multirow{2}{*}{Material} &  \multicolumn{2}{c} {SF width (DFT-PFDD)} && \multicolumn{2}{c} { SF width (MD-PFDD (GAP))} && \multicolumn{2}{c} { Standalone MD (ZW)} \\
            \cline{2-3}    \cline{5-6}    \cline{8-9}
    & edge (\AA) & screw (\AA) && edge (\AA) & screw (\AA) &&      edge (\AA) & screw (\AA) \\
        \hline
        \centering
           Cu &  35.9 & 12.8 && 23.1 & 7.9 && 34.03 $\pm$ 2.25 & 15.5 $\pm$ 2.5 \\   
           Cu2Al  & 43.7 & 12.9  && 23.1 & 7.9 && 38.14 $\pm$ 2.36 & 18.17 $\pm$ 2.04 \\   
           Cu4Al  & 51.5 &  15.5 && 25.8 & 7.9  && 52.18 $\pm$ 3.02 & 20.18 $\pm$ 1.41\\   
           Cu8Al  & 67.2 & 18.1  && 28.5 & 7.9  && 64.77 $\pm$ 2.61 & 22.12 $\pm$ 1.6\\   
           Cu10Al  & 72.4 & 18.1  && 33.9 & 7.9  && 75.83 $\pm$ 4.61 & 25.4 $\pm$ 2.7\\ 
        \hline
    \end{tabular}
    \label{tab:sfw}
\end{table}

\end{document}